\documentclass[reprint, amsmath,amssymb,aps,]{revtex4-1}

\usepackage{graphicx}
\usepackage{dcolumn}
\usepackage{bm}
\usepackage{amsmath}
\usepackage{float}
\usepackage{multirow}
\usepackage{scrextend}
\usepackage{hyperref}
\usepackage{cancel}

\begin{document}
\preprint{APS/123-QED}

\title{Three-body decay of $\Lambda_c^{*} (2765)$ and determination of its spin-parity}

\author{A. J. Arifi$^{1,2}$}
\author{H. Nagahiro$^{3,1}$}
\author{A. Hosaka$^{1,2}$}
\author{K. Tanida$^{2}$}
\affiliation{ 
$^1$Research Center for Nuclear Physics (RCNP), Osaka University, Ibaraki, Osaka 567-0047, Japan\\
$^2$Advanced Science Research Center, Japan Atomic Energy Agency, Tokai, Ibaraki 319-1195, Japan\\
$^3$Department of Physics, Nara Women's University, Nara 630-8506, Japan}

\date{\today}

\begin{abstract}
We study three-body decays of $\Lambda_c^{*}(2765) \to \Lambda_c^+\pi^+\pi^-$ by using effective Lagrangians in a non-relativistic framework.
We consider the sequential decays through $\Sigma_c(2455)\pi$ and $\Sigma_c^*(2520)\pi$ in intermediate states which are dominant contributions.
The coupling constants in the effective Lagrangians are computed in the quark model.
We demonstrate that the ratio $R= \Gamma(\Lambda_c^*\to\Sigma_c^*(2520)\pi)/\Gamma(\Lambda_c^*\to \Sigma_c(2455)\pi)$ and angular correlations are sensitive to the spin and parity of $\Lambda_c^{*}(2765)$.
Thus, the measurement of these observables  in experimental facilities such as Belle and LHCb can provide useful constraints to determine the spin and parity of $\Lambda_c^{*}(2765)$.
\end{abstract}

\pacs{Valid PACS appear here}
\maketitle

\section{INTRODUCTION}

In the past decades, several $\Lambda_c^*$ resonances are experimentally observed in the study of their three-body decays into $\Lambda_c^+\pi^+\pi^-$.
The low-lying excited states $\Lambda_c^*(2595)$ and $\Lambda_c^*(2625)$ have been generally accepted as a $p$-wave doublet in Particle Data Group (PDG)~\cite{Tanabashi:2018oca}. 
The quark model and other calculations give the consistent results to each others as $p$-wave states with $\lambda$-mode excitations, {\it e.g.} see references~\cite{Yoshida:2015tia,Nagahiro:2016nsx,Zhong:2007gp}.
Their three-body decays have been investigated in detail in our previous studies~\cite{Arifi:2017sac,Arifi:2018yhr}.

In contrast to $\Lambda_c^*(2595)$ and $\Lambda_c^*(2625)$, the information of $\Lambda_c^*(2765)$ or $\Sigma_c^*(2765)$ is still poor experimentally.
The broad $\Lambda_c^*(2765)$ or $\Sigma_c^*(2765)$ resonance was observed by CLEO \cite{Artuso:2000xy} in $\Lambda_c^+\pi^+\pi^-$ final state and later by Belle \cite{Abe:2006rz}.
In PDG, this state still has a one-star rating with unknown spin and parity~\cite{Tanabashi:2018oca}.
However, the experimental study is underway~\cite{Joo:2014fka}.
Recently, the isospin has been determined to be $I=0$ by Belle \cite{Abdesselam:2019bfp}.
Therefore, this resonance should be written as $\Lambda_c^*(2765)$.

The mass spectrum of charmed baryons has been studied intensively in various theoretical models~\cite{Copley:1979wj, Capstick:1986bm, Oh:1995ey, Ebert:2007nw, Valcarce:2008dr, Chen:2009tm, Ebert:2011kk, Chen:2014nyo, Yoshida:2015tia, Shah:2016nxi, Lu:2016ctt, Kumakawa:2017ffl,Gandhi:2019xfw}.
The decay pattern of $\Lambda_c^*(2765)$ has also been investigated theoretically.
In particular, its two-body decay of $\Lambda_c^*(2765) \to \Sigma_c^{(*)}\pi$ has been discussed in various models~\cite{Zhong:2007gp, Cheng:2015naa, Nagahiro:2016nsx, Chen:2016iyi,Guo:2019ytq}.
However, a complication lies in that the resonance is considered to be in the region of excitation energy of $2\hbar \omega$, where many configurations with different spins and parities are possible.

An interesting feature of this resonance is its excitation energy of about 500 MeV.  
In fact, there exist baryon resonances systematically in various flavor contents of $u,d,s$ quarks with similar excitation energy, 
known as the Roper resonance for the nucleon sector~\cite{Roper:1964zza}, with the spin and parity $1/2^+$.  
The excitation energy 500 MeV is significantly lower than the amount that is expected by the quark model.
This fact has brought many ideas such as collective monopole vibration~\cite{Brown:1983ib}, strong coupling with meson clouds~\cite{Suzuki:2009nj}, the band head of rotational states of a deformed state~\cite{Takayama:1999kc} and so forth.  
If the same feature is also seen for charmed baryons, the flavor-independent nature will provide an interersting aspect of QCD dynamics for hadron resonances.

\begin{figure}[b]
\vspace{-0.8cm}
\centering
\includegraphics[scale=1]{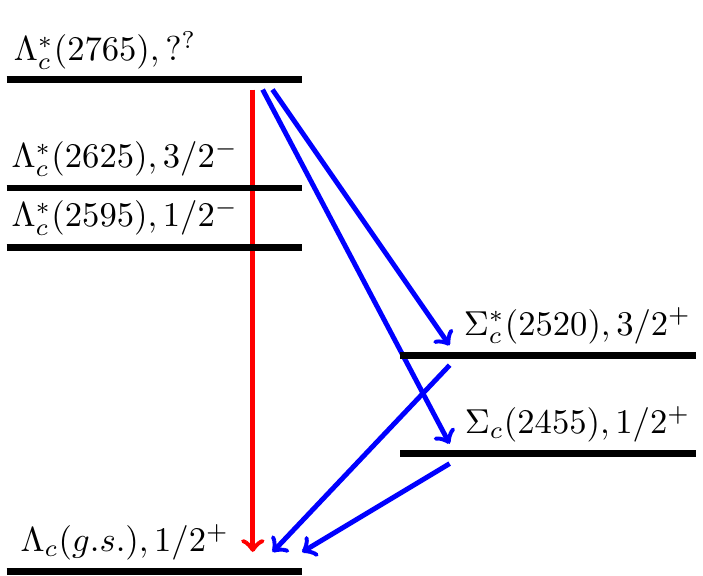}
\caption{\label{level} Three-body decay of $\Lambda_c^{*}(2765)$ into $\Lambda_c^+\pi^+\pi^-$. Blue arrows represent sequential processes, and a red arrow corresponds to the direct process.}
\end{figure}

In the present paper, we aim to study three-body decays of $\Lambda_c^*(2765) \to \Lambda_c^+ \pi^+\pi^-$ as shown in Fig.~\ref{level} using their Dalitz plots and other related quantities.  
We show that different assignments of spin and parity for $\Lambda_c^*(2765)$ clearly differentiate them, the comparison of which with experimental data will be useful for the determination of its spin and parity.  

The essential ingredients are the elementary three-particle vertices for such as $\Lambda_c^* \Sigma_c \pi$.  
They form the so-called sequential decay processes, which are known to be dominant in the present decays.  
In Ref.~\cite{Nagahiro:2016nsx}, some of such vertices for $\Lambda_c^*(2765)$ with possible spins and parities have been studied in the quark model.
In the present work, we complete the calculations for all possible states up to the $2 \hbar \omega$ region in the quark model.  
They include states of spin and parity $J^P=1/2^\pm, 3/2^\pm, 5/2^\pm$ and $7/2^+$ with $\lambda$ and $\rho$ mode orbital excitations.
We then compute the three-body decays of $\Lambda_c^*(2765)$ for all these cases by using effective Lagrangians.
The resulting Dalitz plots and related quantities turn out to be sensitive to the spin and parity of $\Lambda_c^*(2765)$. 

\begin{figure*}
\centering
\includegraphics[scale=1]{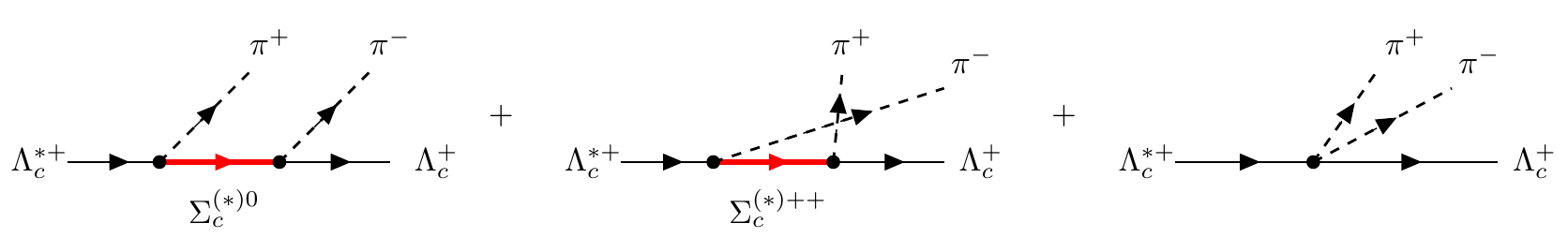}
\caption{\label{3body} Feynman diagrams of three-body decays of $\Lambda_c^{*+}$ into $\Lambda_c^+\pi^+\pi^-$. 
  The first two diagrams represent sequential processes going through $\Sigma_c^{(*)0}$ and $\Sigma_c^{(*)++}$. 
  The last diagram corresponds to the direct process.}
\end{figure*}

The rest of the paper is organized as follows.  
In Sec. II, we explain the decay amplitudes by using effective Lagrangians and their coupling constants. 
We also explain the computation of the three-body decay amplitudes and discuss the kinematics.
In Sec. III, we discuss two-body decays of $\Lambda_c^*(2765)$ with various configurations in the quark model.
In Sec. IV, we discuss three-body decays of $\Lambda_c^*(2765)$ with various configurations and analyze their Dalitz plots and other related quantities.
Finally, we give a summary in Sec. V. 

\section{FORMALISM}
The Feynman diagrams of the three-body decay of $\Lambda_c^*(2765)$ are shown in Fig.~\ref{3body} where the left two are the so-called sequential processes, while the most right one is the direct process. 
In the experimental observation~\cite{Abe:2006rz}, it is implied that the direct process is not important.
Accepting this fact, we will focus on the sequential processes going through $\Sigma_c(2455)\pi$ and $\Sigma_c^*(2520)\pi$ in intermediate states.
To compute these sequential decay processes, we introduce effective Lagrangians describing various vertices of the diagrams. 
We perform the calculations in the non-relativistic approximation, which is suitable for the decays of charmed (heavy) baryons.  

\subsection{Two-body decays}

Here we compute two-body decay amplitudes of the first vertex $\Lambda_c^* \to \Sigma_c^{(*)}\pi$
and second vertex $\Sigma_c^{(*)} \to \Lambda_c \pi$ where $\Sigma_c^{(*)}$ is either $\Sigma_c(2455)$ or $\Sigma_c^*(2520)$.
There are two decay processes in the first vertex, (1) $\Lambda_c^* \to \Sigma_c (2455)\pi$ and (2) $\Lambda_c^* \to \Sigma_c^*(2520)\pi$.
As mentioned in introduction, in the present study, we consider $J^P=1/2^\pm, 3/2^\pm, 5/2^\pm$ and $7/2^+$ for $\Lambda_c^*(2765)$.

In the calculation, we denote the spin operators $\boldsymbol{\sigma}$ for spin-1/2 particles and $\boldsymbol{\Sigma}$ for spin-3/2 particles.
We also introduce the spin transition operators ${\bf S}$ for transitions from spin 3/2 to 1/2, ${\bf T}$ for those from spin 5/2 to 3/2, and  ${\bf U}$ for those from spin 7/2 to 5/2.
These operators form  scalar products with the pion momentum ${\bf p}$ at the vertices. 
Moreover, we introduce $V_{ij}$ for transitions from spin 3/2 to 3/2 with a $d$-wave pion, $W_{ijk}$ and $X_{ijk}$ for those from spin 3/2 to 3/2 and spin 5/2 to 3/2, respectively, with an $f$-wave pion.
These spin transition operators are represented in the Cartesian basis.  
They are related to those in the spherical basis that are given by the Clebsh-Gordan coefficients,
\begin{eqnarray}
\left\langle J_f\ m_f \left| S^L_\mu \right| J_i\ m_i \right\rangle & = & \left(J_i \ m_i\ L\ \mu \bigl| J_f\ m_f \right), \label{matrix_el}
\end{eqnarray}
where the rank of the operator $L$ follows the partial wave of the pion.  
For example, $V_{ij}$ is such an operator of rank two.
Note that in defining Eq.~(\ref{matrix_el}), we set the reduced matrix element unity except for the  $\boldsymbol{\sigma}$ and  $\boldsymbol{\Sigma}$ spin matrices.
The arbitrariness of it is absorbed into the coupling constants.

With those ingredients, for $\Lambda_c^* (1/2^-)$, the decay amplitudes are given by
\begin{eqnarray}
-i\mathcal{T}_{\Lambda_c^* \to \Sigma_c \pi} (s) &=& g_1^s\ \chi^\dagger_{\Sigma_c} \chi_{\Lambda_c^*}, \label{2bodyamp1}\\
-i\mathcal{T}_{\Lambda_c^* \to \Sigma_c^* \pi}(d) &=& g_2^d\ \chi^\dagger_{\Sigma_c^*} ({\bf S}^\dagger \cdot {\bf p}) (\boldsymbol{\sigma} \cdot {\bf p})  \chi_{\Lambda_c^*},
\end{eqnarray}
where $\chi_{\Lambda_c^*}$ and $\chi_{\Sigma_c^{(*)}}^\dagger$ are the spin states of $\Lambda_c^*$ and $\Sigma_c^{(*)}$, respectively, and ${\bf p}$ is the pion momentum.
The coupling constants $g_1^s$ and $g_2^d$ correspond to the Yukawa couplings of the first vertex in the sequential process in Fig.~\ref{3body} going to $\Lambda_c^* \to \pi \Sigma_c$ and $\Lambda_c^* \to \pi \Sigma_c^*$ processes, respectively.
The labels ($s$) and ($d$) on the left hand side indicate that the partial waves of $\pi \Sigma_c^{(*)}$ are $s$ and $d$ wave, respectively.
The labels are also shown as superscripts in each coupling constant.

For $\Lambda_c^* (3/2^-)$, the amplitudes are written as
\begin{eqnarray}
-i\mathcal{T}_{\Lambda_c^* \to \Sigma_c \pi} (d)&=& g_1^d\ \chi^\dagger_{\Sigma_c} (\boldsymbol{\sigma} \cdot {\bf p})  ({\bf S} \cdot {\bf p}) \chi_{\Lambda_c^*}, \\
  -i\mathcal{T}_{\Lambda_c^* \to \Sigma_c^* \pi} (s) &=& g_2^s\ \chi^\dagger_{\Sigma_c^*} \chi_{\Lambda_c^*},  \\
  -i\mathcal{T}_{\Lambda_c^* \to \Sigma_c^* \pi} (d)   &=& g_2^d\ \chi^\dagger_{\Sigma_c^*} \left( {\bf p} \cdot {\bf V}\cdot {\bf p} \right) \chi_{\Lambda_c^*},
\end{eqnarray}
where $\Lambda_c^*(3/2^-) \to \Sigma_c^*\pi$ can decay both in $s$ and $d$ waves. 
Accordingly, we define their coupling constants as $g_2^s$ and $g_2^d$, respectively.

For $\Lambda_c^* (5/2^-)$, we have 
\begin{eqnarray}
  -i\mathcal{T}_{\Lambda_c^* \to \Sigma_c \pi}(d) &=& g_1^d\ \chi^\dagger_{\Sigma_c}({\bf S} \cdot {\bf p}) ({\bf T} \cdot {\bf p})  \chi_{\Lambda_c^*},\\
  -i\mathcal{T}_{\Lambda_c^* \to \Sigma_c^* \pi} (d) &=& g_2^d\ \chi^\dagger_{\Sigma_c^*} (\boldsymbol{\Sigma} \cdot {\bf p}) ({\bf T} \cdot {\bf p}) \chi_{\Lambda_c^*}, \label{aa}
\end{eqnarray}
where $\Lambda_c^*(5/2^-) \to \Sigma_c^* \pi$ can decay only in $d$ wave.
We do not consider $g$ wave because it is not possible due to the brown muck selection rule in the quark model, as we will discuss in the next subsection.

For positive parity cases, amplitudes are calculated in a similar way. For $\Lambda_c^* (1/2^+)$, they are given by
\begin{eqnarray}
  -i\mathcal{T}_{\Lambda_c^* \to \Sigma_c \pi} (p)&=& g_1^p\ \chi^\dagger_{\Sigma_c} (\boldsymbol{\sigma} \cdot {\bf p}) \chi_{\Lambda_c^*},\\
  -i\mathcal{T}_{\Lambda_c^* \to \Sigma_c^* \pi} (p)  &=& g_2^p\ \chi^\dagger_{\Sigma_c^*} ({\bf S}^\dagger \cdot {\bf p}) \chi_{\Lambda_c^*}.
\end{eqnarray}

For $\Lambda_c^* (3/2^+)$,
\begin{eqnarray}
  -i\mathcal{T}_{\Lambda_c^* \to \Sigma_c \pi}(p) &=& g_1^p\ \chi^\dagger_{\Sigma_c}({\bf S} \cdot {\bf p})  \chi_{\Lambda_c^*}, \\
  -i\mathcal{T}_{\Lambda_c^* \to \Sigma_c^* \pi}  (p)&=& g_2^p\ \chi^\dagger_{\Sigma_c^*} (\boldsymbol{\Sigma} \cdot {\bf p})  \chi_{\Lambda_c^*},\\
  -i\mathcal{T}_{\Lambda_c^* \to \Sigma_c^* \pi}  (f) &=& g_2^f\ \chi^\dagger_{\Sigma_c^*} \left(  W_{ijk}\ p_{i}\  p_{j}\ p_{k} \right)  \chi_{\Lambda_c^*},
\end{eqnarray}
where $\Lambda_c^*(3/2^+) \to \Sigma_c^*\pi$ can decay both in $p$ and $f$ waves.

For $\Lambda_c^* (5/2^+)$, 
\begin{eqnarray}
  -i\mathcal{T}_{\Lambda_c^* \to \Sigma_c \pi}(f) &=& g_1^f\ \chi^\dagger_{\Sigma_c} (\boldsymbol{\sigma} \cdot {\bf p}) ({\bf S} \cdot {\bf p}) ({\bf T} \cdot {\bf p})  \chi_{\Lambda_c^*}, \quad\quad \\
  -i\mathcal{T}_{\Lambda_c^* \to \Sigma_c^* \pi} (p)  &=& g_2^p\ \chi^\dagger_{\Sigma_c^*} ({\bf T} \cdot {\bf p})  \chi_{\Lambda_c^*},\\
  -i\mathcal{T}_{\Lambda_c^* \to \Sigma_c^* \pi} (f)  &=& g_2^f\ \chi^\dagger_{\Sigma_c^*}   \left(  X_{ijk}\ p_{i}\  p_{j}\ p_{k} \right)   \chi_{\Lambda_c^*},
\end{eqnarray}
where $\Lambda_c^*(5/2^+) \to \Sigma_c^*\pi$ can decay both in $p$ and $f$ waves.

For $\Lambda_c^* (7/2^+)$, 
\begin{eqnarray}
  -i\mathcal{T}_{\Lambda_c^* \to \Sigma_c \pi}(f)      &=& g_1^f \chi^\dagger_{\Sigma_c}     ({\bf S} \cdot {\bf p}) ({\bf T} \cdot {\bf p}) ({\bf U} \cdot {\bf p})  \chi_{\Lambda_c^*}, \\
  -i\mathcal{T}_{\Lambda_c^* \to \Sigma_c^* \pi} (f)  &=& g_2^f \chi^\dagger_{\Sigma_c^*}  (\boldsymbol{\Sigma} \cdot {\bf p})({\bf T} \cdot {\bf p})   ({\bf U} \cdot {\bf p})  \chi_{\Lambda_c^*}, \label{bb}\quad\quad
\end{eqnarray}
where the $h$-wave decay is forbidden due to the brown muck selection rule in the quark model.

For the second vertex, we calculate the $\Sigma_c \to \Lambda_c \pi$ and $\Sigma_c^* \to \Lambda_c \pi$ amplitudes as
\begin{eqnarray}
  -i\mathcal{T}_{\Sigma_c \to \Lambda_c \pi} (p) &=& g_3^p\ \chi^\dagger_{\Lambda_c} \left( \boldsymbol{\sigma} \cdot {\bf p}\right)  \chi_{\Sigma_c}, \\
  -i\mathcal{T}_{\Sigma_c^* \to \Lambda_c \pi} (p) &=& g_4^p\ \chi^\dagger_{\Lambda_c} \left( {\bf S} \cdot {\bf p}\right) \chi_{\Sigma_c^*}, \label{2bodyamp2}
\end{eqnarray}
where $g_3^p$ and $g_4^p$ are Yukawa couplings of the second vertex in sequential process corresponding to $\Sigma_c^{(*)} \to \Lambda_c\pi$.

\subsection{Coupling constants by the quark model}

To determine various coupling constants of the effective Lagrangians, we compute helicity amplitudes both in effective Lagrangians and in the quark model.  
Let us start from helicity amplitudes in effective Lagrangians for the second vertex $\Sigma_c^{(*)} \to \Lambda_c \pi$, $A_{1/2}$,
\begin{eqnarray}
-i A_{1/2}(\Sigma_c \to \Lambda_c \pi) &=& g_3^p \left<\tfrac{1}{2},\tfrac{1}{2}\left|(\boldsymbol{\sigma}\cdot {\bf p})\right|\tfrac{1}{2},\tfrac{1}{2}\right>\nonumber \\
  &=& g_3^p\ p \left<\tfrac{1}{2},\tfrac{1}{2}\left|\sigma_z\right|\tfrac{1}{2},\tfrac{1}{2}\right>\nonumber\\
  &=& g_3^p\ p,\\
-i A_{1/2}(\Sigma_c^* \to \Lambda_c \pi) &=& g_4^p \left<\tfrac{1}{2},\tfrac{1}{2}\left|({\bf S}\cdot {\bf p})\right|\tfrac{3}{2},\tfrac{1}{2}\right> \nonumber\\
  &=& g_4^p\ p  \left<\tfrac{1}{2},\tfrac{1}{2}\left| S_z \right|\tfrac{3}{2},\tfrac{1}{2}\right>\nonumber\\ 
  &=& -\sqrt{\frac{2}{3}} g_4^p\  p.
\end{eqnarray}
Because of the spin 1/2 of the final state $\Lambda_c$, only helicity 1/2 is allowed as indicated by the subscript 1/2.
Similarly, the helicity amplitudes for the first vertex are computed as in the following. 
For $\Lambda_c^*(1/2^-)$ decays,
\begin{eqnarray}
  -i A_{1/2}(\Lambda_c^* \to \Sigma_c   \pi) &=& g_1^s \left<\tfrac{1}{2}, \tfrac{1}{2}\right| \left.\tfrac{1}{2},\tfrac{1}{2}\right>= g_1^s,\\ 
  -i A_{1/2}(\Lambda_c^* \to \Sigma_c^* \pi) &=& g_2^d \left< \tfrac{3}{2}, \tfrac{1}{2}\right|  ({\bf S}^\dagger \cdot {\bf p}) (\boldsymbol{\sigma} \cdot {\bf p}) \left| \tfrac{1}{2}, \tfrac{1}{2}\right>\nonumber\\
  &=& \sqrt{\frac{2}{3}} g_2^d\  p^2.
\end{eqnarray}
For $\Lambda_c^*(3/2^-)$ decays, 
\begin{eqnarray}
  -i A_{1/2}( \Lambda_c^* \to \Sigma_c \pi)  &=& g_1^d \left<  \tfrac{1}{2}, \tfrac{1}{2}\right| (\boldsymbol{\sigma} \cdot {\bf p})  ({\bf S} \cdot {\bf p}) \left| \tfrac{3}{2},  \tfrac{1}{2}\right>\nonumber\\
    &=& -\sqrt{\frac{2}{3}} g_1^d\  p^2.
\end{eqnarray}
For $\Sigma_c^*\pi$ channel, because the spin of  $\Sigma_c^*$ is 3/2, there are two helicity amplitudes $A_{1/2}$ and $A_{3/2}$ and two possible partial waves, $s$ and $d$ waves. 
For the $s$-wave case, the amplitudes are written as
\begin{eqnarray}
  -i A_{1/2}( \Lambda_c^* \to \Sigma_c^* \pi) &=& g_2^s \left< \tfrac{3}{2}, \tfrac{1}{2}\right| \left. \tfrac{3}{2}, \tfrac{1}{2}\right> = g_2^s,\\
  -i A_{3/2}( \Lambda_c^* \to \Sigma_c^* \pi) &=& g_2^s \left< \tfrac{3}{2}, \tfrac{3}{2}\right| \left. \tfrac{3}{2}, \tfrac{3}{2}\right> = g_2^s,
\end{eqnarray}
while for $d$-wave, the amplitudes are given by
\begin{eqnarray}
    -i A_{1/2}( \Lambda_c^* \to \Sigma_c^* \pi) &=& g_2^d \left< \tfrac{3}{2}, \tfrac{1}{2}\right| \left( {\bf p} \cdot {\bf V}\cdot {\bf p} \right) \left| \tfrac{3}{2}, \tfrac{1}{2}\right>\nonumber\\
		&=& g_2^d\ p^2 \left< \tfrac{3}{2}, \tfrac{1}{2}\right| V_{zz} \left| \tfrac{3}{2}, \tfrac{1}{2}\right>\nonumber\\
    &=& -\frac{1}{\sqrt{5}}  g_2^d\  p^2,\\
    -i A_{3/2}(\Lambda_c^* \to \Sigma_c^*  \pi) &=& g_2^d \left< \tfrac{3}{2}, \tfrac{3}{2}\right|  \left( {\bf p} \cdot {\bf V}\cdot {\bf p} \right) \left| \tfrac{3}{2}, \tfrac{3}{2}\right>\nonumber\\
    &=&  \frac{1}{\sqrt{5}}  g_2^d\  p^2.
\end{eqnarray}
Other cases can be calculated in similar manners.
The summary of helicity amplitudes in the effective Lagrangians is given in Table~\ref{eff_coupling}.

Now for quark model calculations, following Ref.~\cite{Nagahiro:2016nsx}, baryon wave functions are formed in the heavy quark basis.  
Namely, a diquark which is formed by two light quarks (brown muck) is combined with the one heavy quark to form baryons.  
Therefore, quark model configurations for $\Lambda_c^*$ states are denoted as $\Lambda_c^*(nl_\xi, J(j)^P)$ where $nl$ stand for the node and orbital angular momentum quantum numbers, and $\xi =\lambda, \rho,\lambda\lambda, \rho\rho$ or $ \lambda\rho$ indicate orbital excitations.
Its spin and parity are denoted by $J(j)^P$, in which $j$ corresponds to the total angular momentum of the brown muck.
In the quark model,  we employ the axial-vector type coupling for the interaction between the pion and a light quark inside a charmed baryon as
\begin{eqnarray}
\mathcal{L}_{\pi qq} = \frac{g_A^q}{2f_\pi} \bar{q} \gamma_\mu\gamma_5 \vec{\tau} q \cdot \partial^\mu \vec{\pi} \label{inter}
\end{eqnarray}
where $g^q_A$ is the quark axial vector coupling constant and $f_\pi=93$ MeV is the pion decay constant. 
Helicity amplitudes are computed by sandwiching the $\pi qq$ interaction in Eq.~(\ref{inter}) by baryon wave functions.
Details are found in Ref.~\cite{Nagahiro:2016nsx}, and here we summarize the results. 

To simplify the notations, we define the quantities as following
\begin{eqnarray}
C_0^\lambda &=&i G\frac{E}{m} a_\lambda F(p),\\
C_0^\rho &=&i G\frac{E}{m} a_\rho F(p),\\
C_2^\lambda &=& \frac{i G M}{a_\lambda (2m+M)}\left[ 2 + \frac{E}{2m} \left(1- \frac{M}{2m+M}\right) \right] F(p),\quad\quad\\
C_2^\rho &=& \frac{ i G}{2a_\rho}\left[ 2 + \frac{E}{2m} \left(1- \frac{M}{2m+M}\right) \right] F(p),\\
C_1 &=& G \left[ 2 + \frac{E}{2m} \left(1- \frac{M}{2m+M}\right) \right] F(p),\\
C_1^{\lambda\lambda} &=& G\frac{E}{m}  \left(\frac{M}{2m+M}\right) F(p),\\
C_1^{\rho\rho} &=&G \frac{E}{2m} F(p),\\
C_3^{\lambda\lambda} &=& \frac{ G M^2}{a_\lambda^2(2m+M)^2}\left[ 2 + \frac{E}{2m} \left(1- \frac{M}{2m+M}\right) \right] F(p),\nonumber\\
\\
C_3^{\rho\rho} &=& \frac{G}{4 a_\rho^2}\left[ 2 + \frac{E}{2m} \left(1- \frac{M}{2m+M}\right) \right] F(p),\\
C_3^{\lambda\rho} &=& \frac{G M}{2a_\lambda a_\rho (2m+M)}\left[ 2 + \frac{E}{2m} \left(1- \frac{M}{2m+M}\right) \right] F(p),\nonumber\\
\end{eqnarray}
where $M$ and $m$ are the masses of the heavy and light quarks. 
We denote the constant $G$ as
\begin{eqnarray}
G &=&\frac{g_A^q}{2f_\pi}.
\end{eqnarray} 
The range of the Gaussian wave functions of $\lambda$ and $\rho$ coordinates are denoted by $a_\lambda$ and $a_\rho$, respectively. 
The Gaussian form factor $F(p)$ is given by
\begin{eqnarray}
F(p) = e^{-p^2_\lambda/4a^2_\lambda}e^{-p^2_\rho/4a^2_\rho}.
\end{eqnarray}
The energy and momentum of an emitted pion are denoted by $E$ and $p$. 
Furthermore, the momentum transfer for the $\lambda$ and $\rho$ modes are given by
\begin{eqnarray}
p_\lambda &=& p \left(\frac{M}{2m+M}\right),\\
p_\rho &=& \frac{p}{2}.
\end{eqnarray}

\begin{table}[t]
\caption{ Helicity amplitudes $A_h$ of $\Lambda_c^* \to \Sigma_c^{(*)}\pi$ decays with various spin and parity assignments, and $\Sigma_c^{(*)} \to \Lambda_c \pi$ decays calculated in effective Lagrangians. }
\centering
\begin{ruledtabular}
\begin{tabular}{cccccccc}
Initial state				& $h$	 &  $A_h(\Lambda_c^* \to \Sigma_c \pi)$  	&  \multicolumn{2}{c}{$A_h(\Lambda_c^* \to \Sigma_c^* \pi)$}	\\ \noalign{\vskip 1mm}  \hline \noalign{\vskip 2mm} 
$\Lambda_c^*(1/2^-) $		& 1/2		& $g_1^s$ 						&  $\sqrt{\frac{2}{3}} g_2^d\  p^2$\\	
$\Lambda_c^*(3/2^-) $		& 1/2		& $-\sqrt{\frac{2}{3}} g_1^d\  p^2$ 		& $g_2^s$ 							& $ -\frac{1}{\sqrt{5}} g_2^d\ p^2$\\	
						& 3/2		& 								& $g_2^s$ 							& $ +\frac{1}{\sqrt{5}}  g_2^d\ p^2$\\	
$\Lambda_c^*(5/2^-) $		& 1/2		& $ \sqrt{\frac{2}{5}} g_1^d\  p^2$ 		& $-\sqrt{\frac{3}{5}} g_2^d\  p^2 $ \\	
						& 3/2		& 								& $-\sqrt{\frac{18}{5}} g_2^d\  p^2 $ \\	
						\noalign{\vskip 2mm} \hline \noalign{\vskip 2mm}  
$\Lambda_c^*(1/2^+) $		& 1/2		& $ g_1^p\ p$ 						& $\sqrt{\frac{2}{3}} g_2^p\  p$\\	
$\Lambda_c^*(3/2^+) $		& 1/2		&  $ -\sqrt{\frac{2}{3}} g_1^p\  p$		& $ g_2^p\ p $							& $ - \frac{3}{\sqrt{35}} g_2^f\ p^3$\\	
						& 3/2		& 								& $3 g_2^p\ p $							& $ +\frac{1}{\sqrt{35}}  g_2^f\ p^3$\\					
$\Lambda_c^*(5/2^+) $		& 1/2		& $\sqrt{\frac{2}{5}} g_1^f\ p^3$			& $-\sqrt{\frac{3}{5}} g_2^p\ p$				& $ + \sqrt{\frac{6}{35}}  g_2^f\ p^3$ \\	
						& 3/2		& 								& $-\sqrt{\frac{2}{5}} g_2^p\ p $				& $ - \frac{3}{\sqrt{35}}  g_2^f\ p^3$\\	
$\Lambda_c^*(7/2^+) $		& 1/2		& $-\sqrt{\frac{8}{35}} g_1^f\ p^3$		& $ \sqrt{\frac{12}{35}} g_2^f\ p^3$ \\	
						& 3/2		& 								& $ \sqrt{\frac{12}{7}} g_2^f\ p^3$\\ \noalign{\vskip 2mm} \hline
\noalign{\vskip 1mm}			&		&  $A_h(\Sigma_c^{(*)} \to \Lambda_c \pi)$	\\ \noalign{\vskip 1mm}\hline				
$\Sigma_c (1/2^+)$			& 1/2		&	 $ g_3^p\  p$\\						
$\Sigma_c^* (3/2^+)$		& 1/2		&	 $-\sqrt{\frac{2}{3}} g_4^p\  p$\\ 
\end{tabular}
\label{eff_coupling}
\end{ruledtabular}
\end{table}

We will demonstrate, for instance, the calculation of the coupling constant for the decay $\Lambda_c^*(1P_\lambda, 3/2(1)^-) \to \Sigma_c \pi$
\begin{eqnarray}
-i A^{el}_{1/2} &=& - i A^{qm}_{1/2}, \nonumber\\
 \sqrt{\frac{2}{3}} g_1^d\  p^2 &=& \left( -\frac{1}{3} \right) p^2 C_2^\lambda, \nonumber\\
  g_1^d &=& -\frac{1}{\sqrt{6}} C_2^\lambda.
\end{eqnarray}

For $\Lambda_c^*(1P_\lambda, 3/2(1)^-) \to \Sigma_c^* \pi$, we have two helicity amplitudes with $h=1/2$ and 3/2.
The coupling constants are obtain by using the relations below
\begin{eqnarray}
-i A^{el}_{1/2} &=& - i A^{qm}_{1/2}, \\
-i A^{el}_{3/2} &=& - i A^{qm}_{3/2}.
\end{eqnarray}
For simplicity, we define $(D_s^{1/2}, D_s^{3/2})$ and $(D_d^{1/2}, D_d^{3/2})$ as the coefficients of the momenta $p^0$ and $p^2$, respectively, in the quark model amplitude for helicity 1/2 and 3/2 as shown in superscripts.
Then, we obtain 
\begin{eqnarray}
g_2^s -\frac{1}{\sqrt{5}} g_2^d\ p^2 &=& D_s^{1/2} + D_d^{1/2} p^2,\\
g_2^s + \frac{1}{\sqrt{5}}  g_2^d\ p^2 &=& D_s^{3/2}+ D_d^{3/2} p^2,
\end{eqnarray}
where there are $s$-wave and $d$-wave amplitudes. 
From the equations above, we can determine the coupling constants $g_2^s$ and $g_2^d$ as
\begin{eqnarray}
g_2^s &=& \tfrac{1}{2} (D_s^{1/2}+D_s^{3/2})  + \tfrac{1}{2} (D_d^{1/2}+D_d^{3/2}) p^2 \nonumber \\
	&=& -\frac{1}{\sqrt{2}}  C_0^\lambda +  \frac{1}{3\sqrt{2}} p^2 C_2^\lambda , \\	
g_2^d &=& - \tfrac{\sqrt{5}}{2} (D_d^{1/2} - D_d^{3/2})\nonumber\\
	&=&  -\frac{\sqrt{5}}{3\sqrt{2}} C_2^\lambda.
\end{eqnarray}

Similarly, we can compute other coupling constants.  
One remark is that for some spin and parity $J^P$, one of the possible partial waves in decaying channels is missing due to the selection rule for the brown muck.  
For instance, for the case of $5/2^-$, possible partial waves are $d$ and $g$ waves.  
The transition (pion emission) occurs between the brown mucks of $j^P = 1^-$ in $\Lambda_c^*(5/2^-)$ and of $j^P = 1^+$ in $\Lambda_c(g.s.)$.  
Due to the pion's spin and parity $0^-$, the transition into the $g$ wave is forbidden.  
We can discuss similarly other cases.  
The results of forbidden partial waves are shown in Table~\ref{corr}.
This explains the discussions around Eqs. (\ref{aa}) and (\ref{bb}).  
We tabulate the coupling constants of $\Lambda_c^*$ and $\Sigma_c^*$ in terms of the quark model for various cases in Table~\ref{coupling}.

\renewcommand{\arraystretch}{1.8}

\begin{table}[t]
\caption{ Coupling constants of the effective Lagrangians in terms of the quark model. 
The quark model configurations are denoted as $\Lambda_c^*(nl_\xi, J(j)^P)$,
the meaning of which is defined in the text.}
\centering
\begin{ruledtabular}
\begin{tabular}{ccll}
Excitation										&	Channel			& Coupling constant  \\ \hline
$\Lambda_c^*(1P_\lambda, 1/2(1)^-) $				&$\Sigma_c\pi(s)$ 		& $g_1^s =   -\frac{1}{\sqrt{2}} C_0^\lambda + \frac{1}{3\sqrt{2}}p^2 C_2^\lambda $	\\
											&$\Sigma_c^*\pi(d)$		& $g_2^d = -\frac{1}{\sqrt{6}}  C_2^\lambda$	\\	
$\Lambda_c^*(1P_\lambda, 3/2(1)^-) $				&$\Sigma_c\pi(d)$		& $g_1^d = \frac{1}{\sqrt{6}} C_2^\lambda$	\\	
											&$\Sigma_c^*\pi(s)$		& $g_2^s =  -\frac{1}{\sqrt{2}}  C_0^\lambda + \frac{1}{3\sqrt{2}} p^2 C_2^\lambda $\\	
											&$\Sigma_c^*\pi(d)$		& $g_2^d = -\frac{\sqrt{5}}{3\sqrt{2}}  C_2^\lambda$\\	
$\Lambda_c^*(1P_\rho, 5/2(2)^-) $					&$\Sigma_c\pi(d)$		& $g_1^d = \frac{1}{\sqrt{6}} C_2^\rho$\\	
											&$\Sigma_c^*\pi(d)$		& $g_2^d =  -\frac{1}{3\sqrt{2}} C_2^\rho$\\ 	
$\Lambda_c^*(2S_{\lambda\lambda}, 1/2(0)^+) $		& $\Sigma_c\pi(p)$		& $g_1^p =  \frac{1}{3\sqrt{2}} C_1^{\lambda\lambda} -\frac{1}{6\sqrt{2}}  p^2 C_3^{\lambda\lambda} $\\		
											&$\Sigma_c^*\pi(p)$		& $g_2^p = -\frac{1}{3}\sqrt{\frac{3}{2}}  C_1^{\lambda\lambda}+ \frac{1}{6}\sqrt{\frac{3}{2}} p^2 C_3^{\lambda\lambda} $\\	
$\Lambda_c^*(1D_{\lambda\lambda}, 3/2(2)^+) $		& $\Sigma_c\pi(p)$		& $g_1^p =  - \sqrt{\frac{5}{12}} C_1^{\lambda\lambda} +\sqrt{\frac{1}{60}} p^2 C_3^{\lambda\lambda}  $\\	
											& $\Sigma_c^*\pi(p)$	& $g_2^p = -\frac{1}{6\sqrt{5} } C_1^{\lambda\lambda} +\frac{1}{30\sqrt{5}} p^2 C_3^{\lambda\lambda}$ \\	
											& $\Sigma_c^*\pi(f)$		& $g_2^f  = -\frac{\sqrt{7}}{10}  C_3^{\lambda\lambda}$\\	
$\Lambda_c^*(1D_{\lambda\lambda}, 5/2(2)^+) $		&$\Sigma_c\pi(f)$		& $g_1^f  = \frac{1}{2\sqrt{6}} C_3^{\lambda\lambda} $\\	
											&$\Sigma_c^*\pi(p)$		& $g_2^p =  -\frac{1}{\sqrt{2} } C_1^{\lambda\lambda} +\frac{1}{5\sqrt{2}} p^2 C_3^{\lambda\lambda} $ \\  
											&$\Sigma_c^*\pi(f)$		& $g_2^f  = -\frac{\sqrt{7}}{15} C_3^{\lambda\lambda} $\\  
$\Lambda_c^*(1D_{\lambda\rho}, 7/2(3)^+) $			&$\Sigma_c\pi(f)$		& $g_1^f  = \frac{1}{2\sqrt{3}} C_3^{\lambda\rho}$\\	
											&$\Sigma_c^*\pi(f)$		& $g_2^f  = -\frac{1}{6} C_3^{\lambda\rho}$\\  \noalign{\vskip 2mm}  \hline	
$\Sigma_c (1S,1/2(1)^+)$							&$\Lambda_c\pi(p)$		& $g_3^p = -\frac{1}{\sqrt{3}}  C_1$\\						
$\Sigma_c^* (1S,3/2(1)^+)$						&$\Lambda_c\pi(p)$		& $g_4^p = -C_1$ \\												
\end{tabular}
\label{coupling}
\end{ruledtabular}
\end{table}
\renewcommand{\arraystretch}{1}

\subsection{Three-body decays}

Let us calculate the three-body decay amplitude of $\Lambda_c^{*}(2765) \rightarrow \Lambda_c \pi^+ \pi^-$ for the sequential processes as described in Fig.~\ref{3body}. 
The amplitude of the first Feynman diagram with an intermediate $\Sigma_c$ is expressed by
\begin{eqnarray}
-i\mathcal{T} \left[\Sigma_c^0(1/2^+)\right] = - i \frac{\mathcal{T}_{\Sigma_c^0 \rightarrow \Lambda_c^+\pi^-}\mathcal{T}_{\Lambda_c^{*+} \rightarrow \Sigma_c^0\pi^+} }{m_{23}-m_{\Sigma_c^0}+\frac{i}{2}\Gamma_{\Sigma_c^0}}, \quad \quad \label{amp1}
\end{eqnarray}
while the amplitude of the cross diagram is written as
\begin{eqnarray}
-i\mathcal{T} \left[\Sigma_c^{++}(1/2^+)\right] &=& -i\frac{ \mathcal{T}_{\Sigma_c^{++}\rightarrow \Lambda_c^+\pi^+}\mathcal{T}_{\Lambda_c^{*+} \rightarrow \Sigma_c^{++}\pi^-} }{m_{13}-m_{\Sigma_c^{++}}+\frac{i}{2}\Gamma_{\Sigma_c^{++}}},\label{amp2}\quad\quad
\end{eqnarray}
where the two-body decay amplitudes $\mathcal{T}$ are taken appropriately from Eqs.~$(\ref{2bodyamp1})$-$(\ref{2bodyamp2})$.
We denote $m_{23}$ and $m_{13}$ as the invariant masses of the subsystem of particle (2, 3) and (1, 3), respectively, where the particle numbers 1, 2, 3 correspond to $\pi^+$, $\pi^-$ and $\Lambda_c^+$. 

The amplitude of the sequential process going through $\Sigma_c^*(3/2^+)$ is calculated similarly.
We emphasize that no phase ambiguity exists for the sequential decay amplitudes when we use the quark model for the coupling constants.
The total amplitude is then a coherent sum, 
\begin{eqnarray}
\mathcal{T} &=& \mathcal{T} \left[\Sigma_c^0\right] +\mathcal{T}\left[\Sigma_c^{++}\right] +\mathcal{T}\left[\Sigma_c^{*0}\right]  + \mathcal{T} \left[\Sigma_c^{*++}\right].
\end{eqnarray}

The actual forms of the three-body decay amplitudes for $\Lambda_c^* (1/2^-)$, for example, are given by
\begin{eqnarray}
  \mathcal{T} [\Sigma_c^0]  &=& F_1^s\ \chi^\dagger_{\Lambda_c}  \left( \boldsymbol{\sigma} \cdot {\bf p_2}\right)  \chi_{\Lambda_c^*}, \quad\quad\\
  \mathcal{T} [\Sigma_c^{*0}] &=& F_2^d\ \chi^\dagger_{\Lambda_c}  \left( {\bf S} \cdot {\bf p_2}\right) ({\bf S}^\dagger \cdot {\bf p_1}) (\boldsymbol{\sigma} \cdot {\bf p_1})  \chi_{\Lambda_c^*}, \quad\quad\\
  \mathcal{T} [\Sigma_c^{++}]  &=& F_3^s\ \chi^\dagger_{\Lambda_c}  \left( \boldsymbol{\sigma} \cdot {\bf p_1}\right)  \chi_{\Lambda_c^*}, \quad\quad\\
  \mathcal{T} [\Sigma_c^{*++}] &=& F_4^d\ \chi^\dagger_{\Lambda_c}  \left( {\bf S} \cdot {\bf p_1}\right) ({\bf S}^\dagger \cdot {\bf p_2}) (\boldsymbol{\sigma} \cdot {\bf p_2})  \chi_{\Lambda_c^*}, 
\end{eqnarray}
where the spin states of initial $\Lambda_c^*(2765)$ and ground state $\Lambda_c$ are denoted by $\chi_{\Lambda_c^*}$ and $\chi^\dagger_{\Lambda_c}$, respectively.
The first and second emitted pions are denoted by ${\bf p_1}$ and ${\bf p_2}$, respectively.
The $F_i$ factor contains information about the coupling constants, normalizations, and the Breit-Wigner function, for instance
\begin{eqnarray}
F_1^s = \frac{g_1^s g_3^p\sqrt{2M_{\Lambda_c^*}} \sqrt{2M_{\Lambda_c}} }{m_{23} - m_{\Sigma_c^0}+ i \Gamma_{\Sigma_c^0}/2}
\end{eqnarray}
where the $g_1^s$ and $g_3^p$ are the coupling constants for the $\Lambda_c^* \Sigma_c\pi$ (first vertex) and $\Sigma_c\Lambda_c\pi$ (second vertex) which have been defined in section II.A.
The three-body decay amplitudes for other spins and parities of $\Lambda_c^*(2765)$ can be computed similarly as in Eqs.~(\ref{amp1}) and (\ref{amp2}).

The three-body decay width is calculated as 
\begin{eqnarray}
\Gamma &=& \frac{(2\pi)^4}{2M} \int  \overline{|\mathcal{T}|^2} {\rm d}\Phi_3(P;p_1,p_2,p_3)\nonumber\\
               &=& \frac{1}{(2\pi)^3}\frac{1}{32M^3} \int   \overline{|\mathcal{T}|^2} {\rm d}m^2_{12} {\rm d}m^2_{23},\quad \label{dalitz}
\end{eqnarray}
where ${\rm d}\Phi_3$ is the three-body phase space and $P$ the momentum of $\Lambda_c^*(2765)$.
From Eq. (\ref{dalitz}), we can see that the three-body decay can be described by a two-dimensional plot of invariant masses $m^2_{12}$ and $m^2_{23}$.

\begin{figure}[t]
\centering
\includegraphics[scale=1.2]{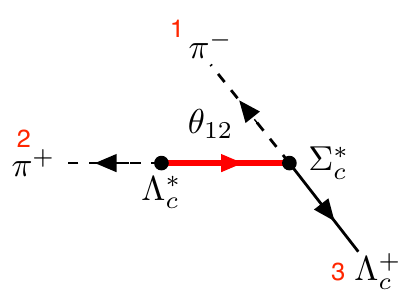}
\caption{\label{correlations} Relative angle of two pions defined in the $\Sigma_c^*$ resonance rest frame.}
\end{figure}

The decay width can also be written as
\begin{eqnarray}
 \Gamma  &=& \frac{1}{8 M^2  (2\pi)^3} \int   \overline{|\mathcal{T}|^2} |{\bf p}_2| |{\bf p}'_1|{\rm d}\cos \theta_{12} {\rm d} m_{23},
\end{eqnarray}
in terms of the invariant mass $m_{23}$ and relative angle of two pions (helicity angle) $\theta_{12}$ as depicted in Fig.~\ref{correlations}.
Here, the momentum ${\bf p}_2$ is calculated in the rest frame of the intermediate $\Sigma_c^{(*)}$ resonance while ${\bf p}_1'$ is calculated in the rest frame of the initial particle $\Lambda_c^*(2765)$.
If we make a plot with a combination of $\cos \theta_{12}$ and $m_{23}$, we will obtain a so-called square Dalitz plot.

For a fixed value of $m_{23}^2$, we can determine the range of $m_{12}^2$ by
\begin{eqnarray}
m_{12}^2 &=&(p_1+p_2)^2 \nonumber\\
	&=& m_1^2 +m_2^2+2 \left( E_1 E_2 - |{{\bf p}}_1| |{{\bf p}}_2| \cos \theta_{12}  \right). \label{1}
\end{eqnarray}
Because the value of $\cos \theta_{12}$ is only between $+1$ and $-1$, the maximum and minimum values of $m_{12}^2$ are
\begin{eqnarray}
(m_{12}^2)_{\pm} &=& m_1^2 +m_2^2+2 \left( E_1 E_2 \pm |{{\bf p}}_1| |{{\bf p}}_2| \right). \label{2}
\end{eqnarray}
We can write the helicity angle in terms of the invariant mass as
\begin{eqnarray}
\cos \theta_{12} = \frac{(m_{12}^2)_{+} + (m_{12}^2)_{-} - 2m_{12}^2}{(m_{12}^2)_{+}- (m_{12}^2)_{-}   }.
\end{eqnarray}
This $\theta_{12}$ angle is used for the study of the angular correlation between the decay products. 
It depends solely on the spin of the participating particles.
In the three-body decay of $\Lambda_c^*(2765)$, those final states and $\Sigma_c^{*}$ intermediate states are known.
Therefore, we can study the spin of the $\Lambda_c^*(2765)$ by analyzing the angular correlations.

\begin{figure}[t]
  \centering
\includegraphics[scale=0.75]{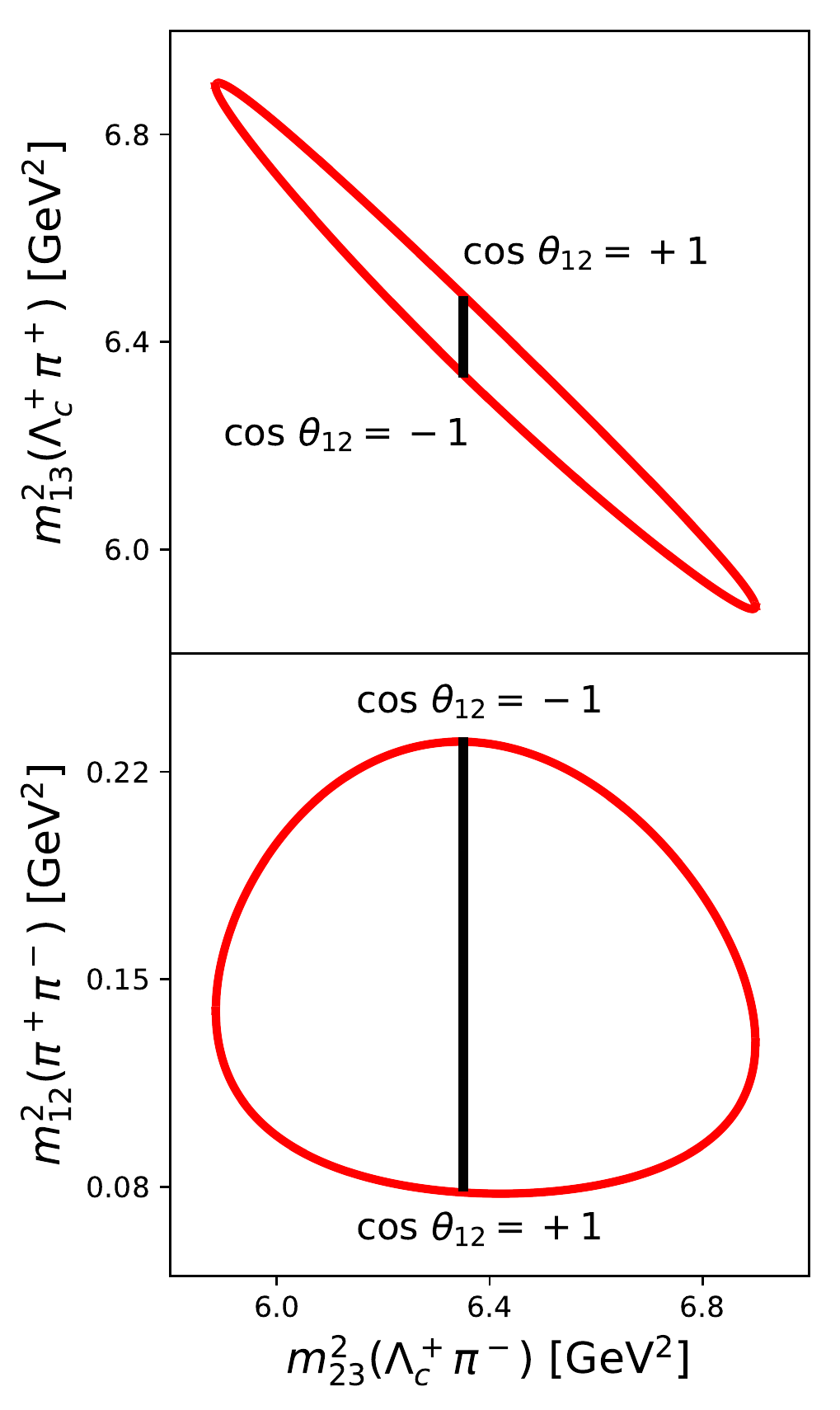}
  \caption{\label{band} Resonance band (vertical solid line) on which angular correlations are studied for different combinations of invariant masses.}
  \end{figure}

The angular correlations are characterized along the resonance bands as depicted in two Dalitz plots with different combinations of invariant masses in Fig.~\ref{band}.
Even though the structures of the two Dalitz plots are essentially the same, the larger area provides a clearer image of the structure on the Dalitz plot, as shown in the lower panel of Fig.~\ref{band}.
We will use the lower one in the following analysis and discussions.

\begin{table*}[t]
\begin{ruledtabular}
\caption{ $\Lambda_c^*(2765) $ decay width into $\Sigma_c(2455)\pi$ and $\Sigma_c^*(2520)\pi$ calculated in the quark model (in unit of MeV). 
$[\Sigma_c^{*}\pi]^{+}$ denotes the isospin summed width by using the isospin-averaged masses.
The quark model configurations are denoted as $\Lambda_c^*(nl_\xi, J(j)^P)$, the meaning of which is defined in the text.
For the mixed $\lambda\rho$ mode, we also show the total angular momentum $\vec{l}=\vec{l}_\lambda + \vec{l}_\rho$ as a subscript $l$ in $J(j)^P_l$.
The ratio is defined by $R= \Gamma(\Lambda_c^*\to \Sigma_c^*\pi)/\Gamma(\Lambda_c^*\to \Sigma_c\pi)$.
We add a subscript $HQ$ in $R_{HQ}$ for the ratio calculated from the heavy-quark symmetry. 
 }
\centering
\begin{tabular}{lccccc}
Excitations								& $[\Sigma_c^{(*)}\pi]_{\text{total}}$ & $[\Sigma_c\pi]^{+}$  & $[\Sigma_c^{*}\pi]^{+}$ & $R$ & $R_{HQ}$\\ \hline
\text{1P-wave}	&\\
$\Lambda_c^*(1P_\lambda, 1/2(1)^-)$			&  	 65.1-146 		& 61.2-140	& 3.90-6.10 	& 0.04-0.06		& -\\
$\Lambda_c^*(1P_\lambda, 3/2(1)^-)$			&  	 52.2-104 		& 7.9-11.9		& 44.3-92.4 	& 5.60-7.80 		& -\\
$\Lambda_c^*(1P_\rho, 1/2(0)^-)$				&  	 - 			&  -			& - 			& -				& -\\
$\Lambda_c^*(1P_\rho, 1/2(1)^-)$				&  	 326-676		& 324-673		& 2.10-3.00 	& 0.004-0.006 		& -\\
$\Lambda_c^*(1P_\rho, 3/2(1)^-)$				&  	 210-413		& 4.20-5.80	& 206-408		& 49.0-70.0			& -\\
$\Lambda_c^*(1P_\rho, 3/2(2)^-)$				&  	9.40-13.1 		& 7.60-10.5	& 1.90-2.70 	& 0.25-0.26 		& 0.22\\	
$\Lambda_c^*(1P_\rho, 5/2(2)^-)$				&	6.30-8.80 		& 3.40-4.70	& 2.90-4.20	& 0.87-0.90 		& 0.76\\	\hline						
\text{2S-wave}	&\\
$\Lambda_c^*(2S_{\lambda\lambda}, 1/2(0)^+)$	& 	1.60-4.50 		& 0.86-2.49	& 0.78-1.98 	& 0.79-0.91 		&  0.80\\ 
$\Lambda_c^*(2S_{\rho\rho}, 1/2(0)^+)$			&	4.69-11.2 	         & 2.60-6.55 	& 2.09-4.60	& 0.70-0.80 		&  0.80\\ \hline
\text{1D-wave}	&\\
$\Lambda_c^*(1D_{\lambda\lambda}, 3/2(2)^+)$	&  	4.70-10.9 		& 4.40-10.1	& 0.33-0.72	 & 0.07-0.08 		& 0.07\\
$\Lambda_c^*(1D_{\lambda\lambda}, 5/2(2)^+)$	&  	1.90-4.40 		& 0.13-0.32	& 1.77-4.04 	& 12.8-13.8		& - \\
$\Lambda_c^*(1D_{\rho\rho}, 3/2(2)^+)$			& 	11.5-23.3 	         & 10.7-21.8	& 0.77-1.43 	& 0.07-0.06		& 0.07\\ 
$\Lambda_c^*(1D_{\rho\rho}, 5/2(2)^+)$			&	4.45-8.63		& 0.13-0.31	& 4.32-8.32 	& 26.8-33.2		& - \\  \hline
\text{1D-wave (mixed)}						\\	
$\Lambda_c^*(1D_{\lambda\rho}, 1/2(1)^+_0)$		&  	 5.47-13.4 	& 4.53-11.3	& 0.93-2.10 	& 0.19-0.21		& 0.20\\
$\Lambda_c^*(1D_{\lambda\rho}, 3/2(1)^+_0)$		&  	 3.47-8.06 	& 1.13-2.82	& 2.33-5.24 	& 1.86-2.06		& 1.99\\	
$\Lambda_c^*(1D_{\lambda\rho}, 1/2(0)^+_1)$		&  	 0.66-1.79 	& 0.42-1.12	& 0.25-0.67 	& 0.60-0.60		& 0.80\\
$\Lambda_c^*(1D_{\lambda\rho}, 1/2(1)^+_1)$		&  	 0.24-0.64 	& 0.21-0.56	& 0.03-0.08 	& 0.15-0.15		& 0.20\\			
$\Lambda_c^*(1D_{\lambda\rho}, 3/2(1)^+_1)$		&  	 0.13-0.35 	& 0.05-0.14	& 0.08-0.21 	& 1.49-1.51		& 1.99\\									
$\Lambda_c^*(1D_{\lambda\rho}, 3/2(2)^+_1)$		&  	 0.28-0.74 	& 0.26-0.70	& 0.02-0.04 	& 0.06-0.06		& 0.07\\			
$\Lambda_c^*(1D_{\lambda\rho}, 5/2(2)^+_1)$		&  	 0.09-0.25		& 0.00		& 0.09-0.25 	& $\infty$			& - \\			
$\Lambda_c^*(1D_{\lambda\rho}, 1/2(1)^+_2)$		&  	 11.4-23.8 	& 9.78-20.5	& 1.61-3.32 	& 0.16-0.16		& 0.20\\			
$\Lambda_c^*(1D_{\lambda\rho}, 3/2(1)^+_2)$		&  	 6.48-13.4 	& 2.45-5.13	& 4.03-8.31 	& 1.62-1.65		& 1.99\\						
$\Lambda_c^*(1D_{\lambda\rho}, 3/2(2)^+_2)$		&  	 23.5-49.3 	& 22.0-46.2	& 1.49-3.11 	& 0.07-0.07		& 0.07\\			
$\Lambda_c^*(1D_{\lambda\rho}, 5/2(2)^+_2)$		&  	 8.92-18.4 	& 0.19-0.40	& 8.73-18.0 	& 44.7-44.9		& - \\				
$\Lambda_c^*(1D_{\lambda\rho}, 5/2(3)^+_2)$		&  	 0.25-0.54 	& 0.22-0.46	& 0.04-0.08 	& 0.17-0.18		& 0.15\\ 	
$\Lambda_c^*(1D_{\lambda\rho}, 7/2(3)^+_2)$		&  	 0.17-0.37		& 0.12-0.26	& 0.05-0.11 	& 0.41-0.43		& 0.35 \\		
\end{tabular}
\label{result_qm}
\vspace{-0.1cm}
\end{ruledtabular}
\end{table*}

\section{ Results for two-body decays}

Let us first revisit two-body decays of $\Lambda_c^*(2765)$ with all possible quark model configurations up to $2\hbar \omega$.
In Table~\ref{result_qm}, we summarize total and partial decay widths for decaying to $\Sigma_c\pi$ and $\Sigma_c^{*}\pi$, and the ratio $R$ which is defined by 
\begin{eqnarray}
R= \frac{\Gamma(\Lambda_c^*(2765)\to \Sigma_c^*(2520)\pi)}{\Gamma(\Lambda_c^*(2765)\to \Sigma_c(2455)\pi)}, \label{ratios}
\end{eqnarray}
for various quark model configurations of $\Lambda_c^*(2765)$.
The uncertainties in the decay widths are from the ambiguities in the quark model parameters such as quark masses and spring constants.

\subsection{Ratios of decay widths}

Model calculations, such as in the quark model, often contain ambiguities in absolute values, which are, however, canceled out by taking the ratios in Eq.~(\ref{ratios}).
This is one of the advantages of studying the ratios.

The ratio $R$ can also be calculated by using the heavy-quark symmetry in a model-independent way~\cite{Isgur:1991wq}.
They provide a measure of how the quark model results follow the heavy-quark symmetry.
Let us consider the decay of $J(j) \to J'(j')+\pi$. 
The initial and final spin of charmed baryon with their corresponding brown muck spin are denoted by $J (j)$ and $J'(j')$, respectively.
In the heavy-quark limit, the heavy quark acts as a static quark, and its spin is decoupled from the light quarks.
Moreover, the decay occurs between the brown muck $j \to j' + \pi$.
As a result, the decay width is computed by using six-$j$ symbols as~\cite{Cheng:2006dk}
\begin{eqnarray}
\Gamma = (2 j + 1) (2J'+1) \left| \left\{ \begin{array}{ccc} J & J' & L\\  j'  & j & s_q \end{array}\right\} \right|^2 p^{(2L+1)} |M_{L}|^2 \quad \quad \label{hqs}
\end{eqnarray}
where $s_q=1/2$ is the heavy-quark spin, $L$ is the relative angular momentum of the final states $\Sigma_c^{(*)}\pi$, $p$ the emitted pion momentum, and $M_L$ the reduced matrix element.
Equation $(\ref{hqs})$ implies that there is a model-independent relation between the decay widths for different $J'$ with the same partial wave. 

For the case of $\Lambda_c^*(1/2^+)$, we have six possible configurations, as in Table~\ref{result_qm}. 
The ratios with different spin $j$ are, for example, given by
\begin{eqnarray}
R [\Lambda_c^*(2S_{\lambda\lambda},1/2(0)^+)]  &=& 0.79-0.91, \\
R [\Lambda_c^*(1D_{\lambda\rho},1/2(1)^+_0)]    &=& 0.19-0.21.
\end{eqnarray}
The calculated ratio for $\Lambda_c^*(1/2^+)$ with $j=0$ is larger than that of $j=1$ by a factor 4.
This factor can be explained by the heavy-quark symmetry using Eq.~(\ref{hqs}). In fact, for $\Lambda_c^*(1/2^+)$ with $j=0$ and $j=1$, the ratios are obtained as 
\begin{eqnarray}
R_{HQ}[\Lambda_c^*(1/2(0)^+)] &=& 2 \times \frac{p(\Sigma_c^*\pi)^3}{p(\Sigma_c\pi)^3} = 0.80, \label{hq1}\\
R_{HQ}[\Lambda_c^*(1/2(1)^+)] &=& \frac{1}{2} \times \frac{p(\Sigma_c^*\pi)^3}{p(\Sigma_c\pi)^3}= 0.20. \quad \label{hq2}
\end{eqnarray}
The ratios for various configurations of $\Lambda_c^*(1/2^+)$ with the same $j$ have similar values.
For instance,  $\Lambda_c^*(2S_{\lambda\lambda},1/2(0)^+)$ and $\Lambda_c^*(2S_{\rho\rho},1/2(0)^+)$ with the same $j=0$ have similar ratios as shown in Table~\ref{result_qm}. 
Consequently, those configurations are difficult to be differentiated by comparing the ratio.

For the case of $\Lambda_c^*(1/2^-)$, there are also two possibilities with $j=0$ and $j=1$.
The decay of $\Lambda_c^*(1/2^-)$ with $j=0$ is forbidden due to the brown muck selection rule, which is indicated by $``-"$ in Table~\ref{result_qm}.
In the quark model, the ratios for $\Lambda_c^*(1/2^-)$ are given by
\begin{eqnarray}
R [\Lambda_c^*(1P_\rho,1/2(0)^-)] 		&=& -,\\
R [\Lambda_c^*(1P_\lambda,1/2(1)^-)] 	&=& 0.04-0.06,\\
R [\Lambda_c^*(1P_\rho,1/2(1)^-)] 		&=& 0.004-0.006.
\end{eqnarray}
The ratio for $\Lambda_c^*(1/2^-)$ with $j=1$ is one order magnitude smaller than for $\Lambda_c^*(1/2^+)$.
This is because $\Lambda_c^*(1/2^-)$ decays into $\Sigma_c^*\pi$ in $d$ wave resulting in a suppression in the ratio as,
\begin{eqnarray}
R [\Lambda_c^*(1/2(1)^-)] = \frac{\Gamma(\Sigma_c^*\pi)_d }{\Gamma(\Sigma_c\pi)_s } \ll 1.
\end{eqnarray}
The ratio is estimated to be much smaller than unity due to $d$-wave nature of $\Sigma_c^*\pi$ decay channel.
In this case, the ratio can not be calculated by the heavy-quark symmetry because the partial waves are different, and therefore the value of $R_{HQ}$ is indicated by $``-"$ in Table~\ref{result_qm}.

For the case of $\Lambda_c^*(3/2^+)$, the ratios calculated in the quark model with $j=1$ and $j=2$, for example, are given by
\begin{eqnarray}
R [\Lambda_c^*(1D_{\lambda\rho},3/2(1)^+_0)] 	&=& 1.86-2.06,\\
R [\Lambda_c^*(1D_{\lambda\lambda},3/2(2)^+)] 	&=& 0.07-0.08.
\end{eqnarray}
The large difference here is understood by the heavy-quark symmetry.
For $\Lambda_c^*\to \Sigma_c^*\pi$ decay, there are two possible partial waves, $p$ wave and $f$ wave.
If we neglect the $f$ wave, we can calculate the ratio for $\Lambda_c^*(3/2^+)$ decays in $p$ wave by the heavy-quark symmetry as
\begin{eqnarray}
R_{HQ}[\Lambda_c^*(3/2(1)^+)] &=& 5 \times \frac{p(\Sigma_c^*\pi)^3}{p(\Sigma_c\pi)^3} = 1.99,  \label{32m1}\\
R_{HQ}[\Lambda_c^*(3/2(2)^+)] &=& \frac{1}{5} \times \frac{p(\Sigma_c^*\pi)^3}{p(\Sigma_c\pi)^3}= 0.07. \quad \label{32m2}
\end{eqnarray}
The results are similar to the quark model calculation.

For $\Lambda_c^*(3/2^-)$, the ratios in the quark model are obtained as
\begin{eqnarray}
R [\Lambda_c^*(1P_\lambda,3/2(1)^-)] 	&=& 5.60-7.80, \\
R [\Lambda_c^*(1P_\rho,3/2(1)^-)] 		&=& 49.0-70.0, \\
R [\Lambda_c^*(1P_\rho,3/2(2)^-)] 		&=& 0.25-0.26.\label{hqs3}
\end{eqnarray}
For $j=1$, the ratio is much larger than unity because the $s$ wave is allowed for the decay into $\Sigma_c^*\pi$ while not for that into $\Sigma_c\pi$,
\begin{eqnarray}
R[\Lambda_c^*(3/2(1)^-)] = \frac{\Gamma(\Sigma_c^*\pi)_s +\Gamma(\Sigma_c^*\pi)_d }{\Gamma(\Sigma_c\pi)_d} \gg 1.
\end{eqnarray}
For the brown muck spin $j=2$, the $s$-wave decay is not allowed due to brown muck selection rule.
Since both channels allow $d$-wave decay, the ratio from the heavy-quark symmetry can be computed as
\begin{eqnarray}
R_{HQ}[\Lambda_c^*(3/2(2)^-)] &=&1 \times \frac{p(\Sigma_c^*\pi)^5}{p(\Sigma_c\pi)^5} = 0.22,
\end{eqnarray}
which is consistent with the quark model in Eq.~(\ref{hqs3}).

For the case of $\Lambda_c^*(5/2^+)$ with $j=2$ and $j=3$, the ratios are calculated as
\begin{eqnarray}
R [\Lambda_c^*(1D_{\lambda\lambda},5/2(2)^+)] 	&=& 12.8-13.8, \\
R [\Lambda_c^*(1D_{\rho\rho},5/2(2)^+)] 			&=& 26.8-33.2, \\
R [\Lambda_c^*(1D_{\lambda\rho},5/2(2)^+_2)] 	&=& 44.7-44.9,\\
R [\Lambda_c^*(1D_{\lambda\rho},5/2(2)^+_1)] 	&=& \infty,\\
R [\Lambda_c^*(1D_{\lambda\rho},5/2(3)^+_2)] 	&=& 0.17-0.18.\label{hqs4}
\end{eqnarray}
For $\Lambda_c^*(1D_{\lambda\rho},5/2(2)^+_1)$, the matrix element of the $\Sigma_c\pi$ decaying channel becomes zero (and hence the ratio becomes infinity) due to conservation of orbital angular momenta.
For $j=2$, the ratio is much larger than unity because the $p$ wave is allowed for the decay into $\Sigma_c^*\pi$ while not for that into $\Sigma_c\pi$,
\begin{eqnarray}
R[\Lambda_c^*(5/2(2)^+)] = \frac{\Gamma(\Sigma_c^*\pi)_p +\Gamma(\Sigma_c^*\pi)_f }{\Gamma(\Sigma_c\pi)_f } \gg 1.
\end{eqnarray}
For $j=3$, $p$ wave is forbidden and only $f$ wave is allowed for both $\Sigma_c\pi$ and $\Sigma_c^*\pi$ decay channels. 
Then, the ratio from the heavy-quark symmetry can be computed as
\begin{eqnarray}
R_{HQ} [\Lambda_c^*(5/2(3)^+)] &=&\frac{5}{4} \times \frac{p(\Sigma_c^*\pi)^7}{p(\Sigma_c\pi)^7}= 0.15,
\end{eqnarray}
which is consistent with the one calculated in the quark model as in Eq.~(\ref{hqs4}).

For the case of $\Lambda_c^*(5/2^-)$, there is only one configuration for the first orbital excitation in the quark model with $j=2$,
\begin{eqnarray}
R [\Lambda_c^*(1P_\rho,5/2(2)^-)] 		&=& 0.87-0.90.
\end{eqnarray}
In this case, only $d$ wave is possible for both decaying channels, the ratio for $\Lambda_c^*(5/2(2)^-)$ is obtained by the heavy-quark symmetry as
\begin{eqnarray}
R_{HQ} [\Lambda_c^*(5/2(2)^-)] &=&\frac{7}{2} \times \frac{p(\Sigma_c^*\pi)^5}{p(\Sigma_c\pi)^5}= 0.76.
\end{eqnarray}

For completeness, we consider $\Lambda_c^*(7/2^+)$ in which it is found as a $1D$-wave state with mixed $\lambda\rho$ mode in the quark model.
The ratio is given by
\begin{eqnarray}
R [\Lambda_c^*(1D_{\lambda\rho},7/2(3)^+_2)] 	&=& 0.41-0.43.
\end{eqnarray}
The ratio for $\Lambda_c^*(7/2(3)^+)$ is computed in the heavy-quark limit for $f$ wave as
\begin{eqnarray}
R_{HQ} [\Lambda_c^*(7/2(3)^+)]	 &=& 3 \times \frac{p(\Sigma_c^*\pi)^7}{p(\Sigma_c\pi)^7}= 0.35.
\end{eqnarray}
The ratio is again consistent with the quark model.

\subsection{Magnitudes of decay widths}

By now, there is only information about the magnitude of $\Lambda_c^*(2765)$ decay width measured by CLEO in the literature.
The measured decay width is about $\Gamma_{\rm exp}\approx 50$ MeV.
As discussed before, the non-resonant contribution is rather small, and the total decay width is dominated by the sequential decays through $\Sigma_c^{(*)}\pi$~\cite{Abe:2006rz}.

As shown in Table~\ref{result_qm}, for negative parity states, $\Lambda_c^*(1P_\lambda, 1/2(1)^-)$ and  $\Lambda_c^*(1P_\rho, 1/2(1)^-)$ gives a rather large decay width due to $s$-wave nature of the decaying channel of $\Sigma_c\pi$.
$\Lambda_c^*(1P_\lambda, 3/2(1)^-)$ and $\Lambda_c^*(1P_\rho, 3/2(1)^-)$ also have a large decay width because of the $s$-wave nature of decaying channel $\Sigma_c^*\pi$.
On the other hand, $\Lambda_c^*(1P_\rho, 3/2(2)^-)$ and $\Lambda_c^*(1P_\rho, 5/2(2)^-)$ give a small decay width due to the $d$-wave nature of decaying channel  $\Sigma_c^{(*)}\pi$.
For positive parity states, almost all configurations give a rather small decay width.
Among various configurations, four cases have a value consistent with data within about factor two.
However, it is fair to say that from the comparison of the total decay widths, one can not determine the spin and parity.
This is the reason that we investigate Dalitz plots together with the angular correlations in the next section.

\begin{figure}[b]
\centering
\includegraphics[scale=0.7]{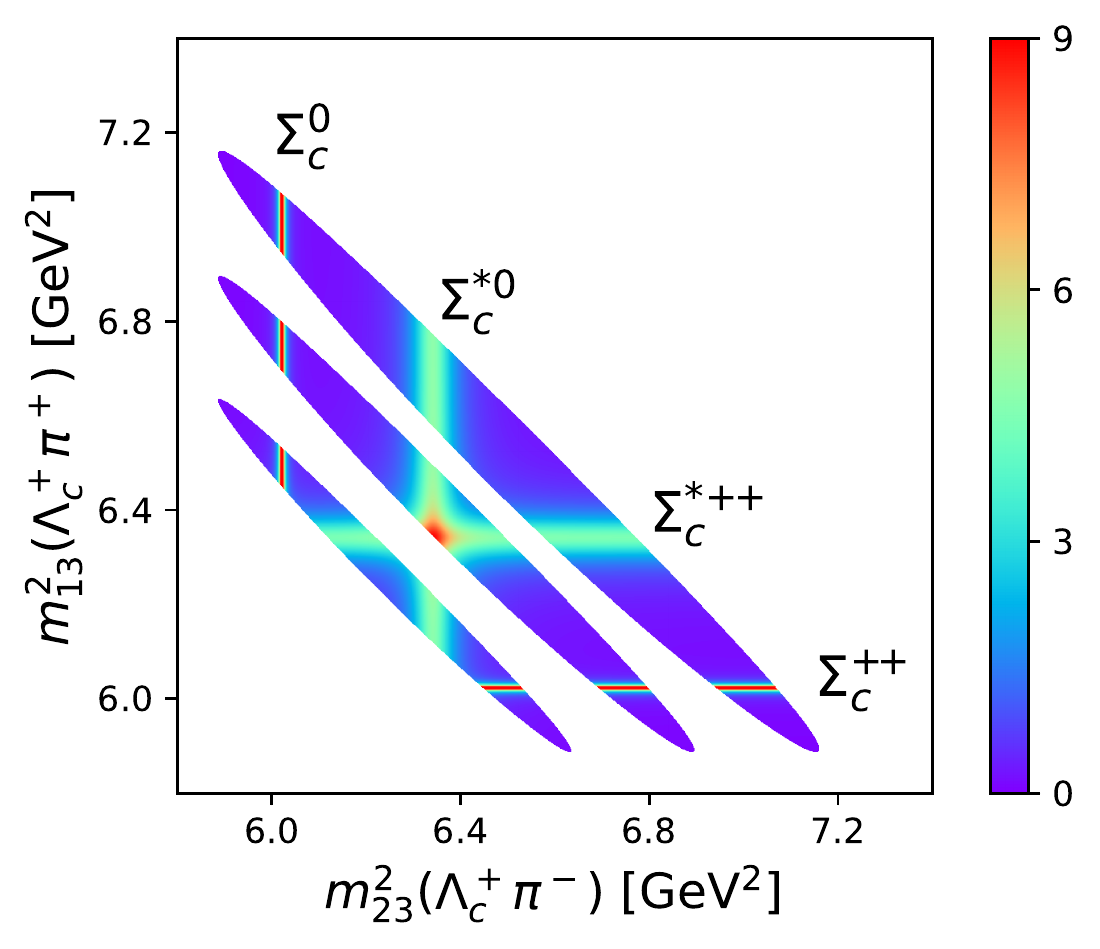}
\caption{\label{convolution} The Dalitz plots of $\Lambda_c^*(2765)$ with mass 2715 (lower), 2765 (middle), and 2815 MeV (upper). 
The bands formed by intermediate states $\Sigma_c^{(*)}$ are indicated in the figure as eye's guides.}
\end{figure}

\begin{figure*}[t]
\centering
\includegraphics[scale=0.52]{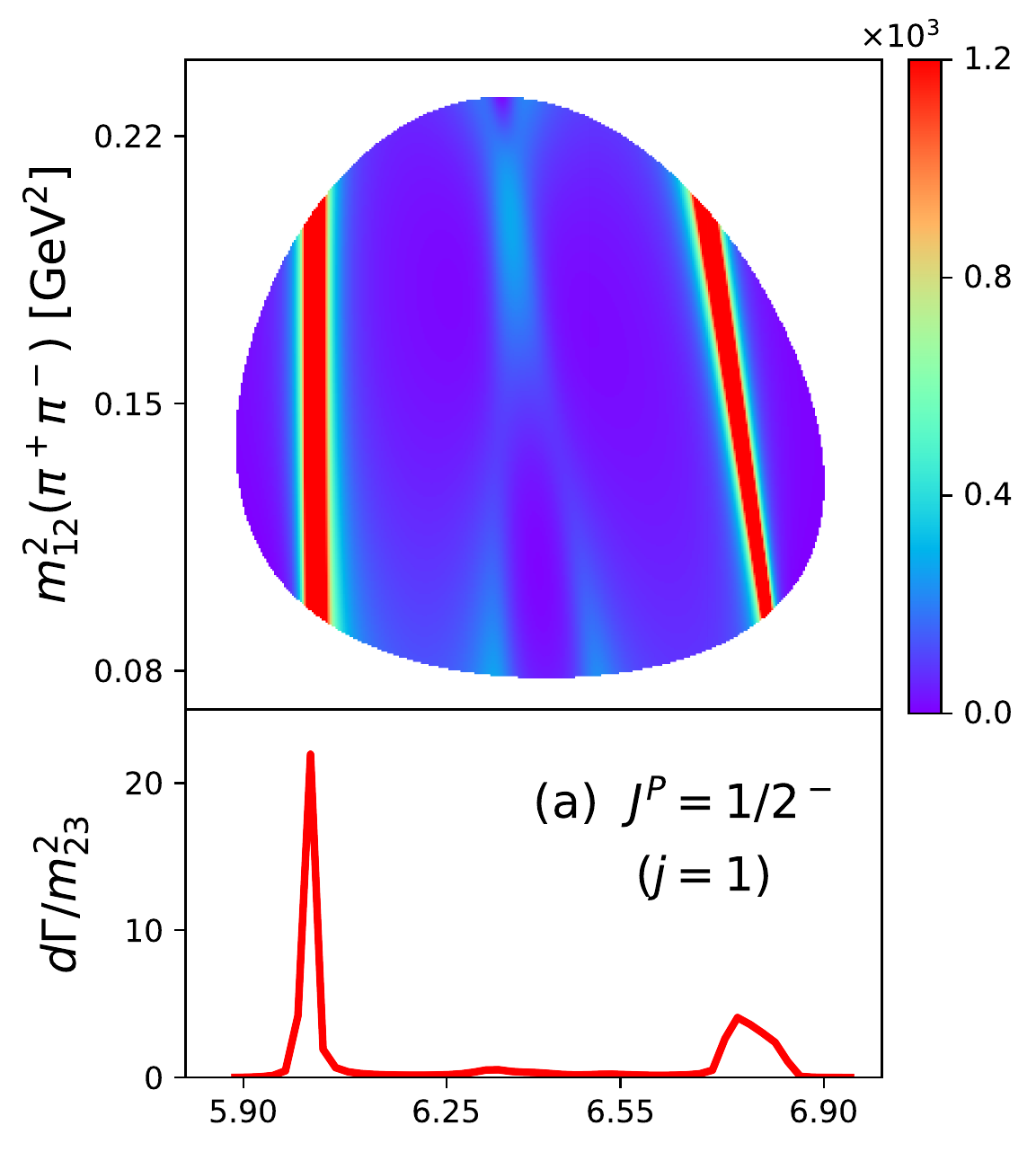}\hspace{0cm}
\includegraphics[scale=0.52]{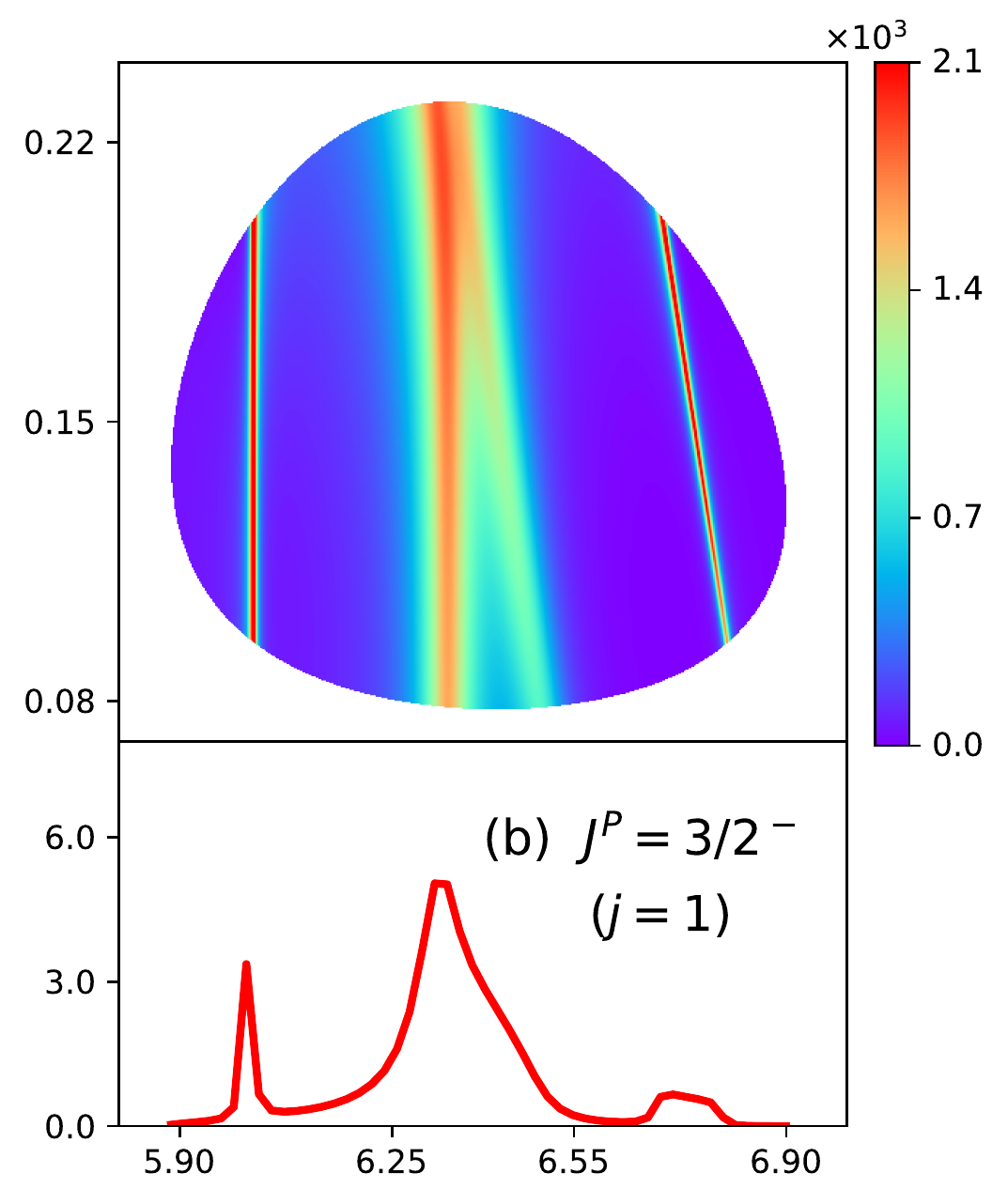}\hspace{0cm}
\includegraphics[scale=0.52]{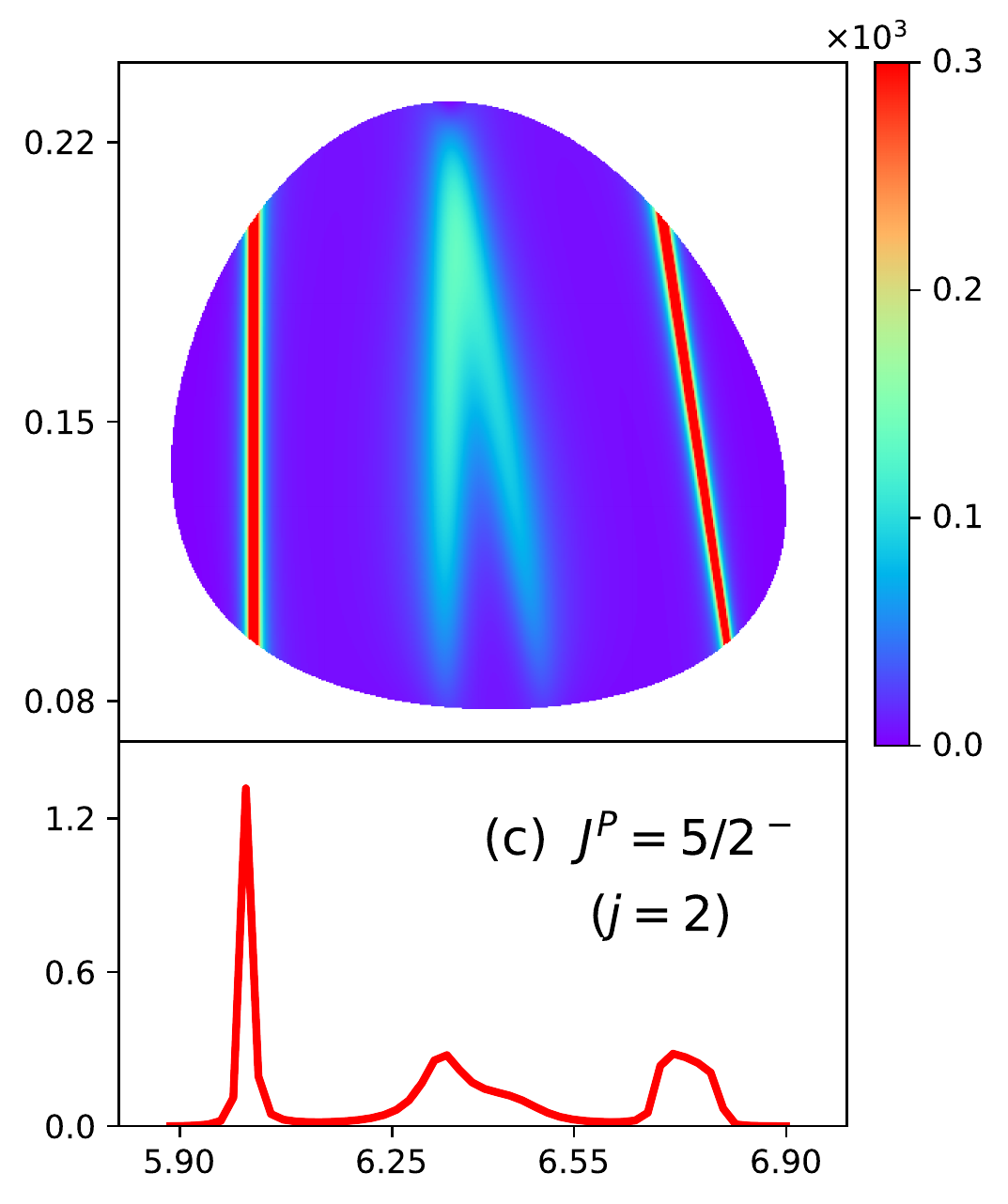}\hspace{0cm}
\includegraphics[scale=0.52]{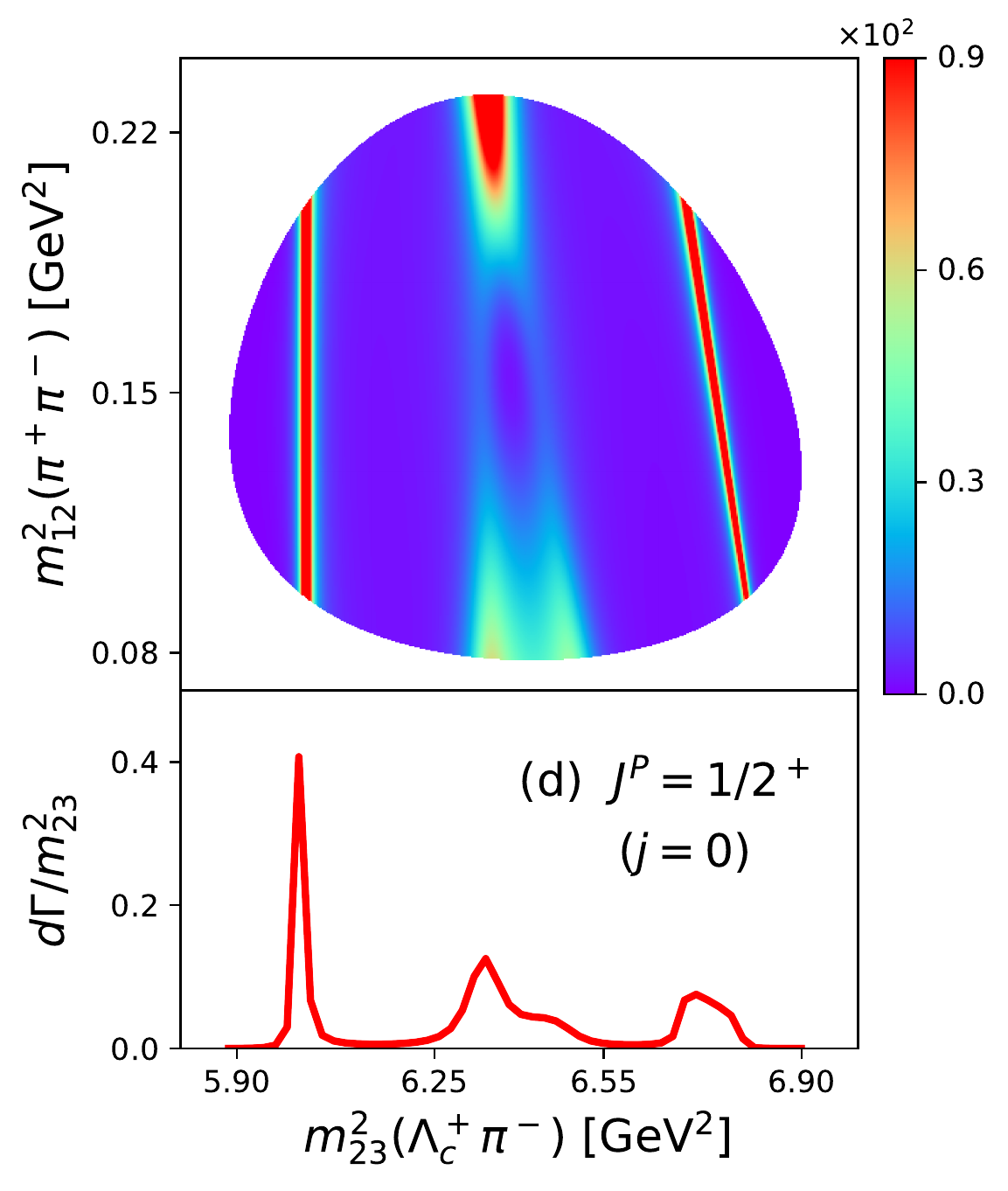}\hspace{0cm}
\includegraphics[scale=0.52]{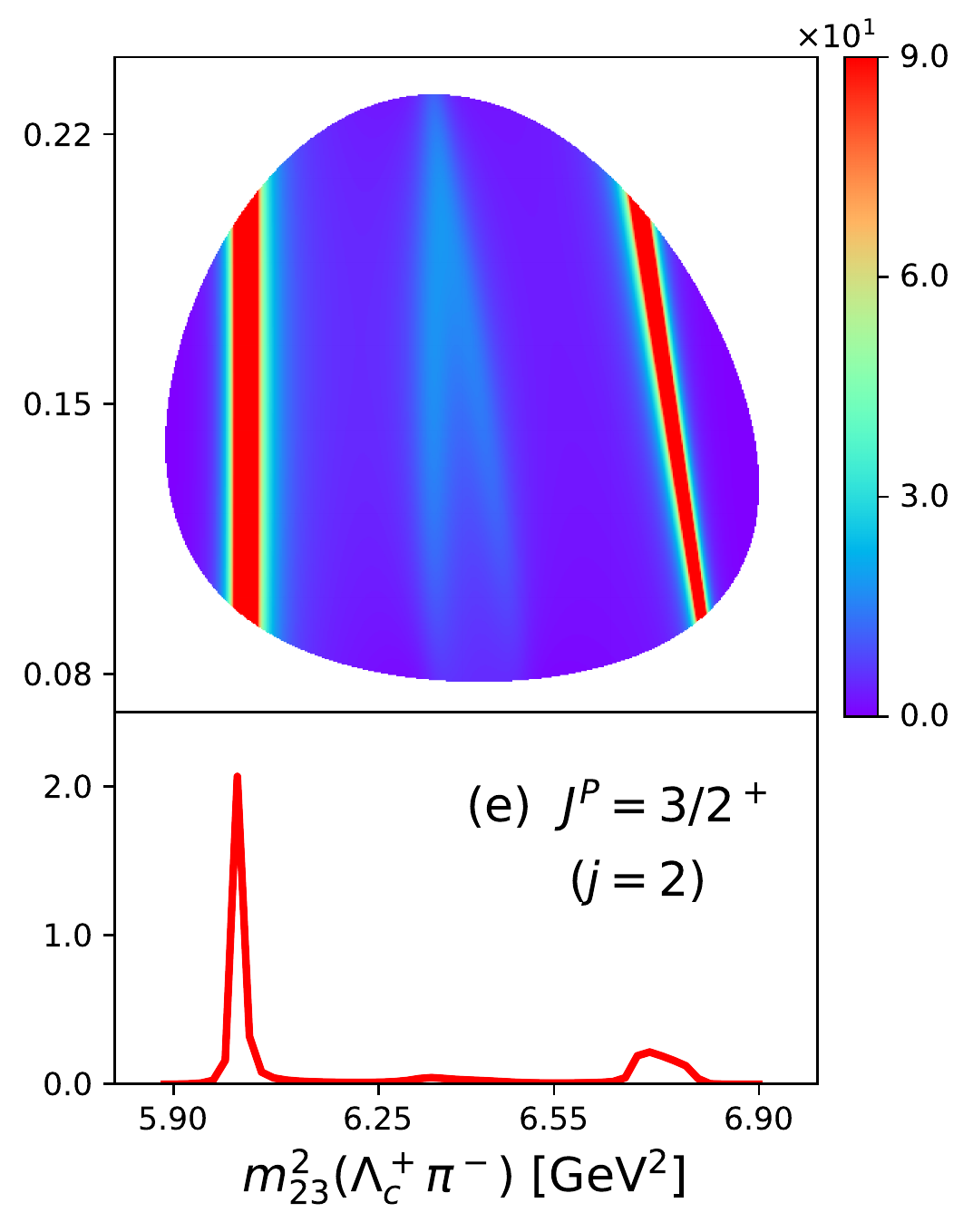}\hspace{0cm}
\includegraphics[scale=0.52]{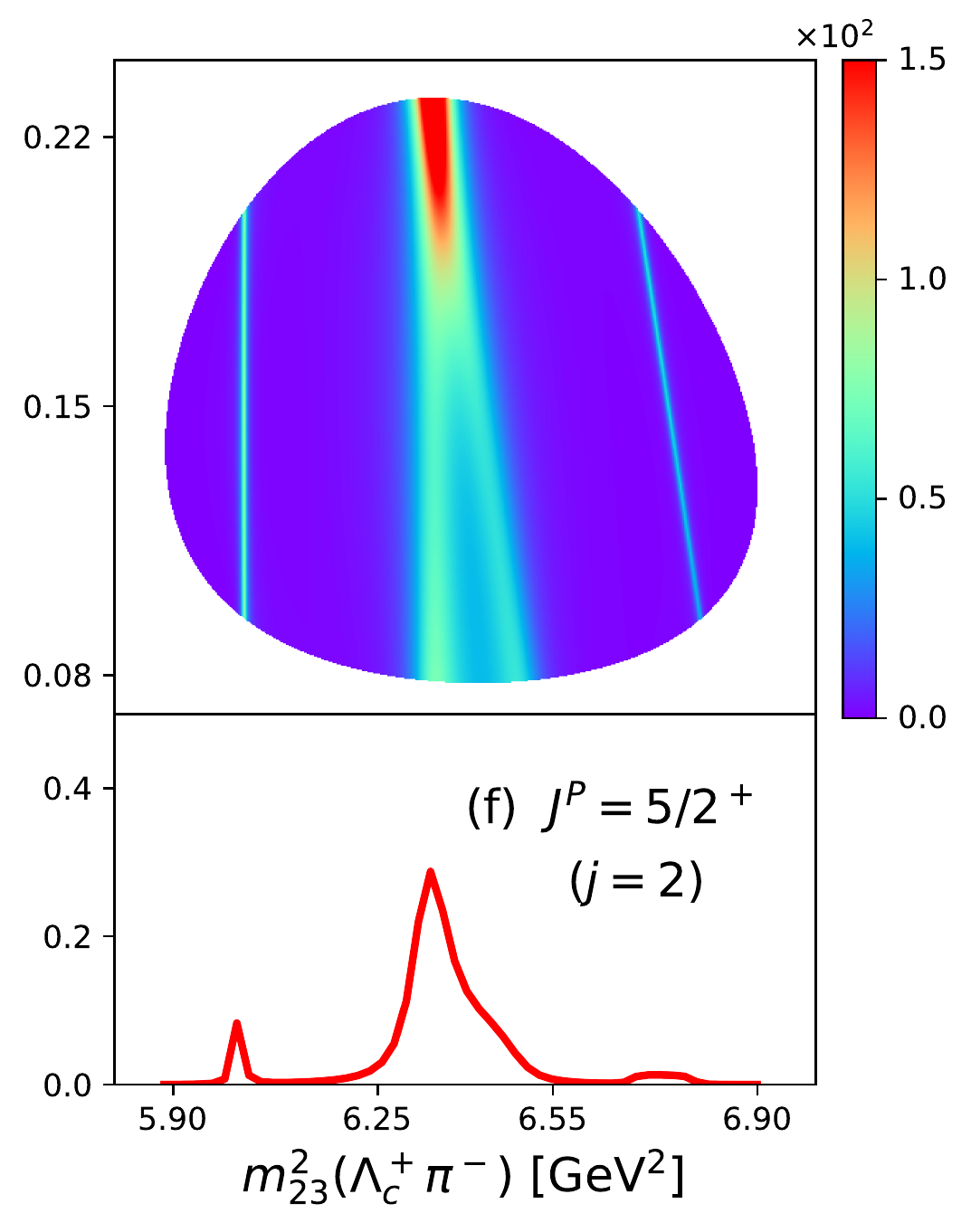}\hspace{0cm}
\caption{\label{resband} The Dalitz plots in the ($m_{23}^2, m_{12}^2)$ plane and the invariant mass plots of $\Lambda_c^+\pi^-$. 
The spins and parities of $\Lambda_c^*(2765)$ as $J^P=1/2^\pm, 3/2^\pm,$ and $5/2^\pm$, along with the corresponding brown muck spin $j$, are indicated in each panel.
The Dalitz plots are made by fixing the initial mass at 2765 MeV.}
\end{figure*}

\section{Results for three-body decays}

Because $\Lambda_c^*(2765)$ is a broad resonance, its mass distributes over a finite width, not in a narrow region.
Consequently, the experimental Dalitz plot may be a superposition of Dalitz plots at various initial masses.
In this paper, we firstly compute various  Dalitz plots at the central value of 2765 MeV in most cases.
Secondly, we will give some remarks as implied by such figures as Fig.~\ref{convolution}, where an example of Dalitz plots are shown for three different masses of $\Lambda_c^*(2765)$. 
Finally, effects of the finite width will be discussed in detail in subsection IV.C.
It turns out that the convoluted Dalitz plots are fairly different from the one computed at a fixed mass.
Therefore, in comparison with actual experimental data, it is important to know whether the data is taken from the mass region distributed over the resonance width or from a fixed (practically within a very narrow energy bin) mass.

\subsection{Dalitz and invariant mass plots}

We investigate all possible spins and parities $J^P = 1/2^\pm, 3/2^\pm, 5/2^\pm$ and $7/2^+$ for $\Lambda_c^*(2765)$.
Among several possible configurations for a given $J^P$, we will select the low-lying configurations of the quark model as follows
\begin{eqnarray}
\Lambda_c^*(1/2^-)  &\to& \Lambda_c^*(1P_\lambda, 1/2(1)^-), \\
\Lambda_c^*(3/2^-)  &\to& \Lambda_c^*(1P_\lambda, 3/2(1)^-), \\
\Lambda_c^* (5/2^-) &\to& \Lambda_c^*(1P_\rho, 5/2(2)^-),\\
\Lambda_c^*(1/2^+) &\to& \Lambda_c^*(2S_{\lambda\lambda}, 1/2(0)^+),\\ 
\Lambda_c^*(3/2^+) &\to& \Lambda_c^*(1D_{\lambda\lambda}, 3/2(2)^+),\\
\Lambda_c^*(5/2^+) &\to& \Lambda_c^*(1D_{\lambda\lambda}, 5/2(2)^+),\\
\Lambda_c^*(7/2^+) &\to& \Lambda_c^*(1D_{\lambda\rho}, 7/2(3)^+).
\end{eqnarray}
For $\Lambda_c^* (5/2^-)$, we select a $\rho$-mode excitation because there is no corresponding $\lambda$ mode.
Note that $\Lambda_c^*(7/2^+)$ appears only as a mixed $\lambda\rho$-mode excitation.
As there are also other configurations for the same spin and parity, we will consider, for example, spin and parity $1/2^+$ and $3/2^+$ with different brown muck spin $j$
\begin{eqnarray}
\Lambda_c^*(1/2^+) &\to& \Lambda_c^*(1D_{\lambda\rho}, 1/2(1)^+),\\ 
\Lambda_c^*(3/2^+) &\to& \Lambda_c^*(1D_{\lambda\rho}, 3/2(1)^+),
\end{eqnarray}
to see the effect of the internal structures.

\begin{figure*}[t]
\centering
\includegraphics[scale=0.52]{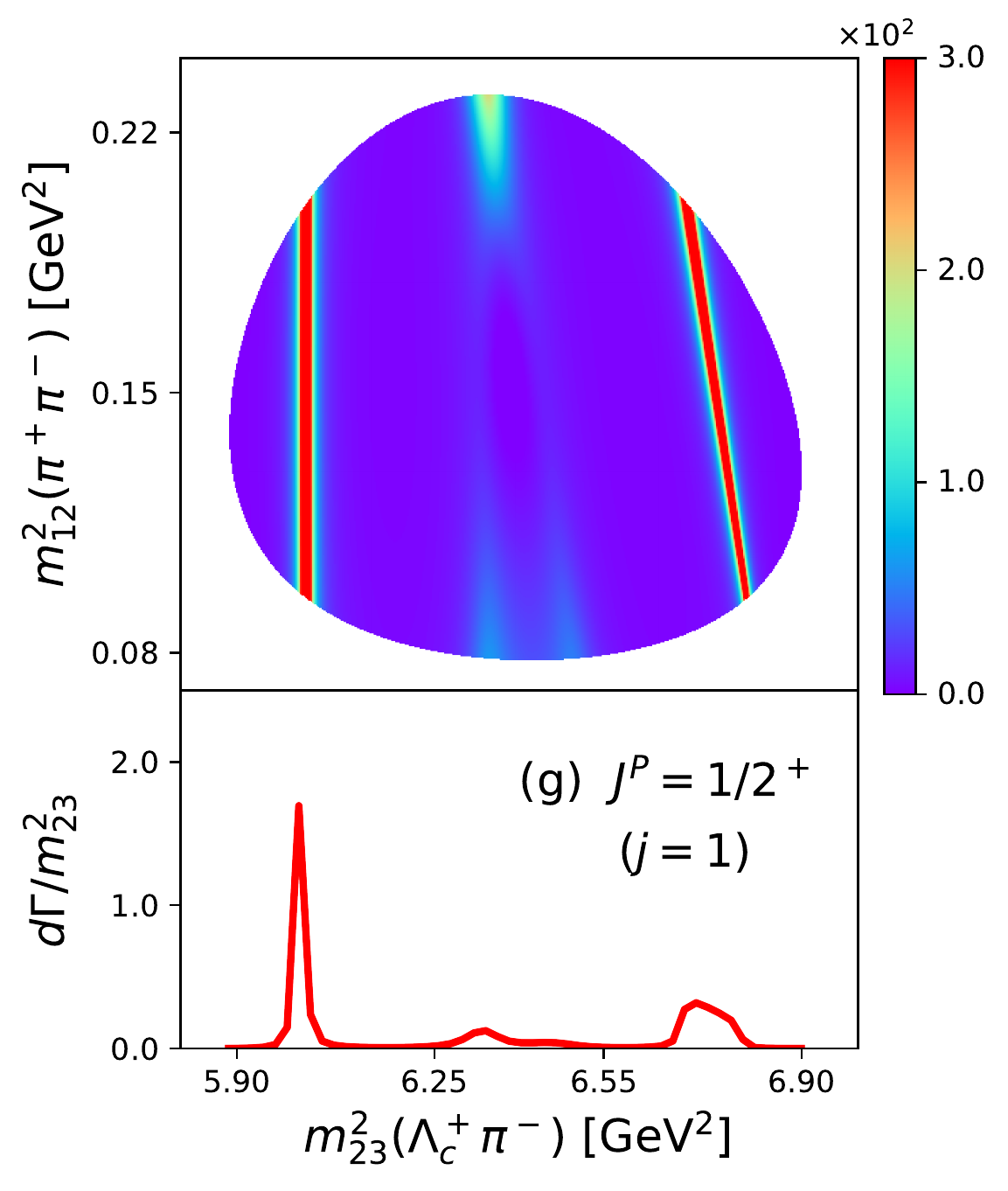}
\includegraphics[scale=0.52]{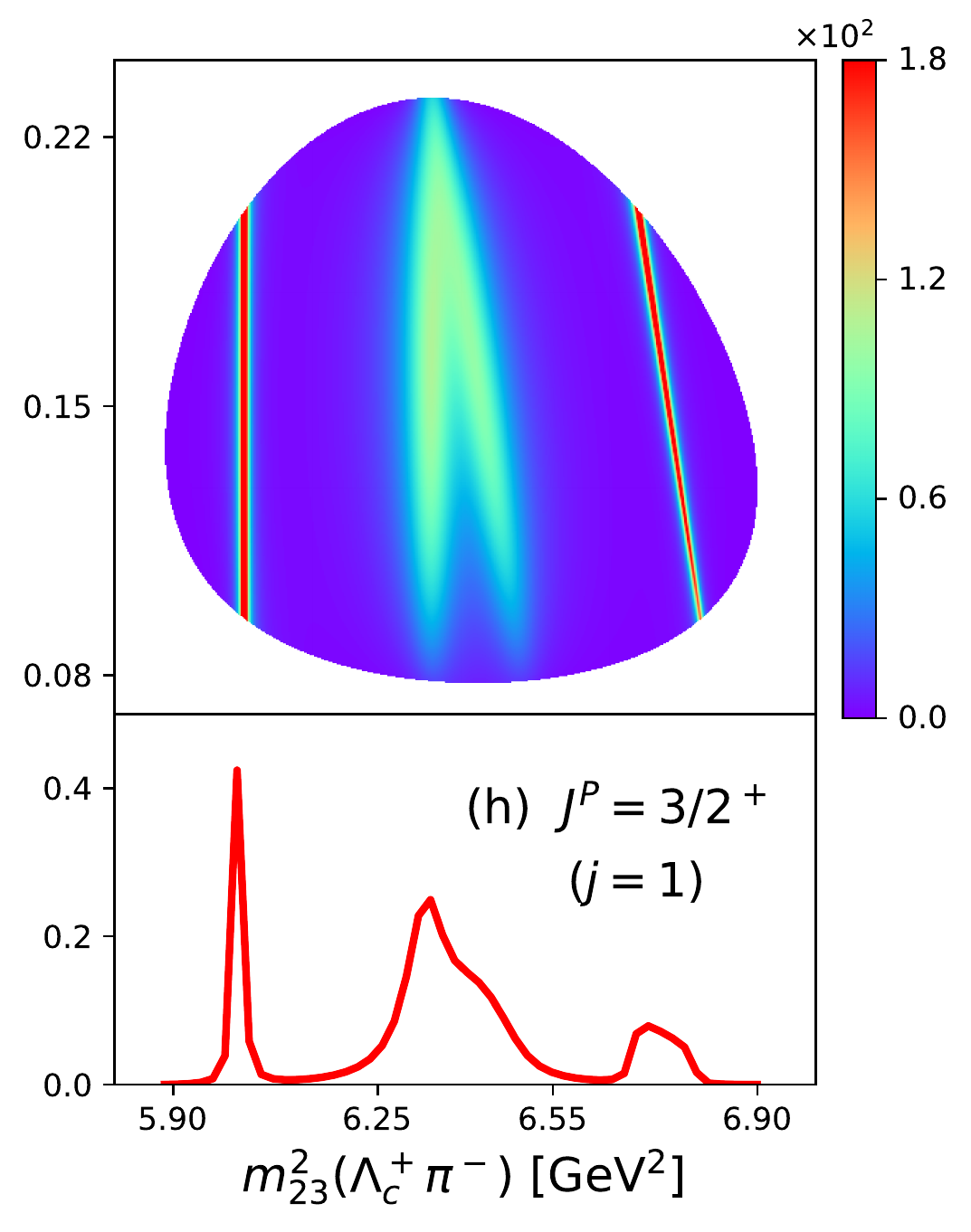}
\includegraphics[scale=0.52]{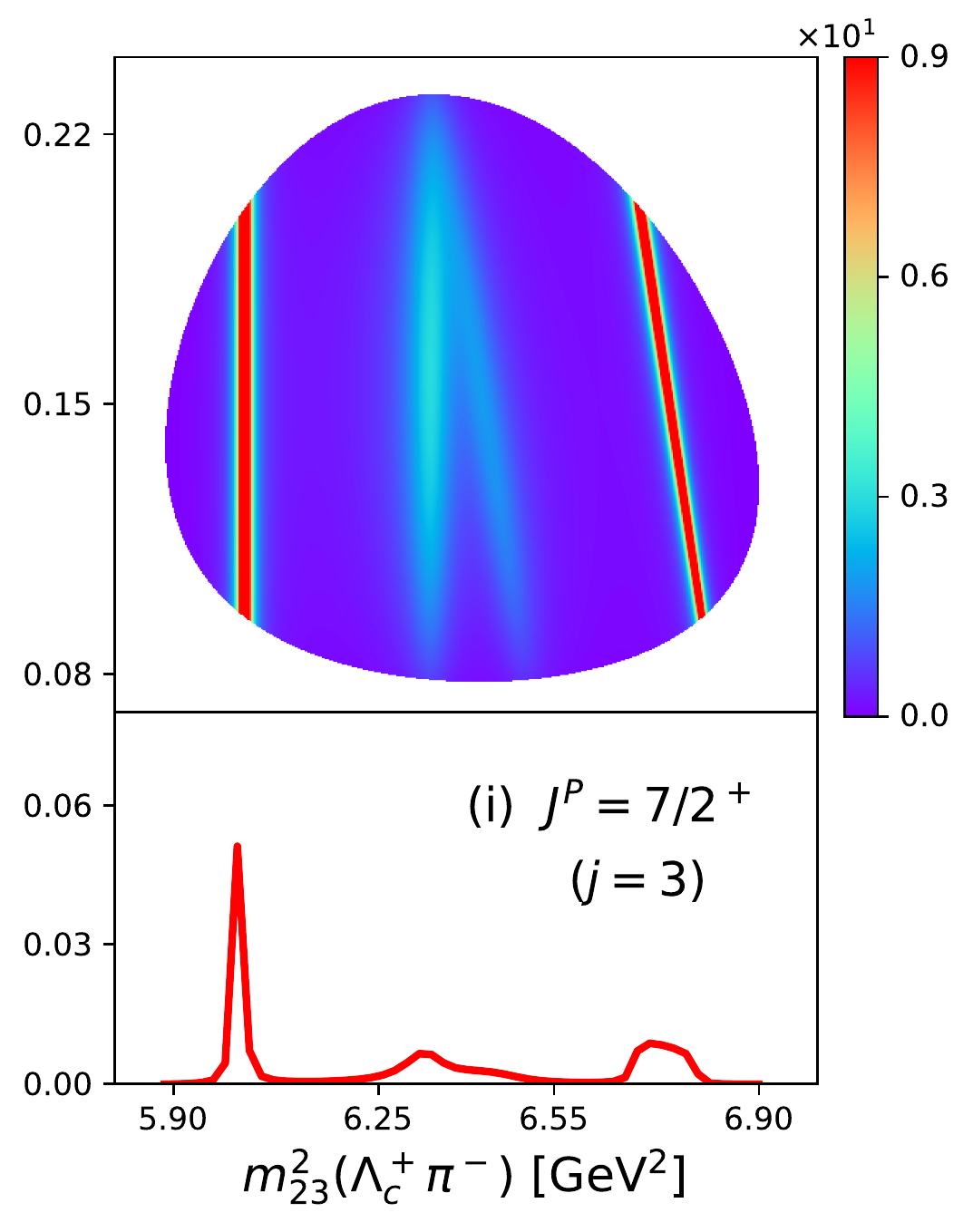}
\caption{\label{12p_lamrho} 
The same as in Fig.~\ref{resband} for the spins and parities of $\Lambda_c^*(2765)$ as $1/2^+$ and $3/2^+$ with different configurations, and $7/2^+$. }
\end{figure*}

The Dalitz plots for various spins and parities are shown in Figs.~\ref{resband} and \ref{12p_lamrho}.
There are four resonance bands in Dalitz plots.
Two resonance bands in the middle correspond to $\Sigma_c^{*0}(2520)$ and $\Sigma_c^{*++}(2520)$,
 while the resonance bands located on the left and right side correspond to $\Sigma_c^0(2455)$ and $\Sigma_c^{++}(2455)$, respectively.
These four resonance bands appear also in the ($m_{23}^2,m_{13}^2$)  plots as shown in Fig.~\ref{convolution}.
In the comparison of these plots, the interference pattern of $\Sigma_c^{*0}$ and $\Sigma_c^{*++}$, and the far-separated location of the two bands for $\Sigma_c^0$ and $\Sigma_c^{++}$ are commonly observed.
Note that the interference occurs only for a specific initial mass around 2765 MeV.
If we choose a higher or lower initial mass such as 2715 or 2815 MeV, the interference disappears as seen from Fig.~\ref{convolution}.

The corresponding invariant mass plots of $\Lambda_c^+\pi^-$ are shown below each Dalitz plot in Fig.~\ref{resband}.
We can also see the corresponding $\Sigma_c^{(*)}$ resonance peaks in the invariant mass plots. 
The peaks on the most left and most right side originating from $\Sigma_c^0$ and $\Sigma_c^{++}$ have different height because the right peak is the kinematical reflection of $\Sigma_c^{++}$ in the $\Lambda_c^+\pi^-$ invariant mass plot.

The Dalitz and invariant mass plots are sensitive to the ratio $R$.
If we look at the ratio for negative parity states of $\Lambda_c^*(2765)$ which are given by
\begin{eqnarray}
R (\Lambda_c^*(1/2^-)) &=& 0.04 - 0.06, \\
R (\Lambda_c^*(3/2^-)) &=& 5.60 - 7.80, \\
R (\Lambda_c^*(5/2^-)) &=& 0.87 - 0.90,
\end{eqnarray}
they are different from each other by one oder of magnitude.
When the ratio is relatively small, the decay process is dominated by the $\Sigma_c$ resonance. 
The $\Sigma_c$ band dominates over the $\Sigma_c^*$ as observed in the Dalitz and invariant mass plot of $\Lambda_c^*(1/2^-)$ decay as shown in Fig.~\ref{resband} (a).
On the contrary, when the ratio is relatively large as in $\Lambda_c^*(3/2^-)$ case, the strong peak of $\Sigma_c^*$ resonance is observed.
Moreover, if the ratio is nearly unity as in $\Lambda_c^*(5/2^-)$, both $\Sigma_c^*$ and $\Sigma_c$ bands appear with equal strength.
These observations also apply to positive parity cases.

In fact, there are several possible quark model configurations for the same spin and parity.
As discussed in the previous section, they differ by the magnitude of the decay width and the ratio $R$. 
Firstly, we have checked that the change of the magnitude will not affect the structure on the Dalitz plot provided that the ratio $R$ remains the same.
Secondly, we investigate other configurations with the same spin and parity, but different $j$, by making other Dalitz plots for $\Lambda_c^*(1/2^+)$ and $\Lambda_c^*(3/2^+)$ with $j=1$ as depicted in Fig.~\ref{12p_lamrho} (g) and (h).
One may notice that the $\Sigma_c^*$ peaks look very different for $\Lambda_c^*(3/2^+)$ with $j=1$ and $j=2$ even though both decaying channels into $\Sigma_c^*\pi$ are $p$ wave.
The difference is governed by the heavy-quark symmetry, as discussed in Eqs.~(\ref{32m1}) and (\ref{32m2}).

\subsection{Angular correlations} 

It has been known that angular correlation (dependence) can help to determine the spin of particles as in gamma-ray spectroscopy in nuclear physics.
A similar analysis can also be applied to hadronic systems.
For instance, the spin 1/2 of $\Sigma_c(2455)$ charmed baryon is determined by analyzing $B^- \to \Lambda_c^+\pi^-\bar{p}$ decay by BaBar~\cite{babar2}.
Since initial $B$-meson has spin 0 and proton has spin 1/2, 
there is helicity conservation such that the $\Sigma_c$ intermediate state in $\Lambda_c\pi$ final state will only have a helicity 1/2 component. 
If $\Sigma_c$'s spin is 1/2, the angular correlation will be flat.
On the contrary, if $\Sigma_c$ has spin 3/2, it will exhibit a concave structure experimentally.
The angular correlation has been found to be flat, confirming that $\Sigma_c(2455)$ has spin 1/2.
A similar analysis can also be done in $\Lambda_c^* \to \Lambda_c\pi\pi$ decay.
Ideally, the angular correlations are determined by the spins of the relevant particles.
In the helicity formalism~\cite{Jacob:1959at}, it is dictated by the Wigner's $D$-functions, which in the present formalism is encoded in the structure of the vertex functions.
The relevant algebra is also done by the tensor formalism~\cite{Zemach:1968zz,Chung:1993da}.

From the Dalitz plots in Fig.~\ref{resband} and \ref{12p_lamrho}, we can observe the angular correlations along the $\Sigma_c(2455)$ look rather flat for all spins and parities of $\Lambda_c^*(2765)$.
This is because only helicity $1/2$ is possible for $\Sigma_c$ resonance which is related to $d^{1/2}_{h_f h_i}(\theta_{12})$ matrix where $h_i$ and $h_f$ are helicities of initial and final states, respectively.
Taking the sum over $h_i$ and $h_f$ for the absolute squared amplitude gives a flat structure in $\theta_{12}$ dependence.

On the other hand, the angular correlations along the $\Sigma_c^*$ resonance bands show characteristic structures through the rank 3/2 $d$-functions, $d^{3/2}_{h_f h_i}(\theta_{12})$.
If $\Lambda_c^*(2765)$'s spin is $1/2$, then the initial helicity takes only $h_i = \pm 1/2$.  
Summing the absolute squared amplitudes over $h_f$ we find the angular correlation $1 +3 \cos^2\theta_{12}$.
If $\Lambda_c^*(2765)$'s spin is $3/2$ or higher, the terms from $h_i = \pm 3/2$ can also contribute.  
Summing the absolute squared amplitude again over $h_f$, we find the correlation  $3\sin^2\theta_{12}$.
In general there are contributions of $h_i = 1/2$ and $3/2$ with a weight of the helicity amplitudes $A_{h_i}(\Lambda_c^* \to \Sigma_c^*\pi)$ for $\Lambda_c^*(2765)$, 
\begin{eqnarray}
W(\theta_{12}) \propto && \left|A_{1/2}(\Lambda_c^* \to \Sigma_c^*\pi)\right|^2 \times (1 +3 \cos^2\theta_{12}) \nonumber\\
		&+& \left|A_{3/2}(\Lambda_c^* \to \Sigma_c^*\pi)\right|^2 \times 3\sin^2\theta_{12}. \quad \quad \label{sin}
\end{eqnarray}

\begin{figure*}[t]
\centering
\includegraphics[scale=0.6]{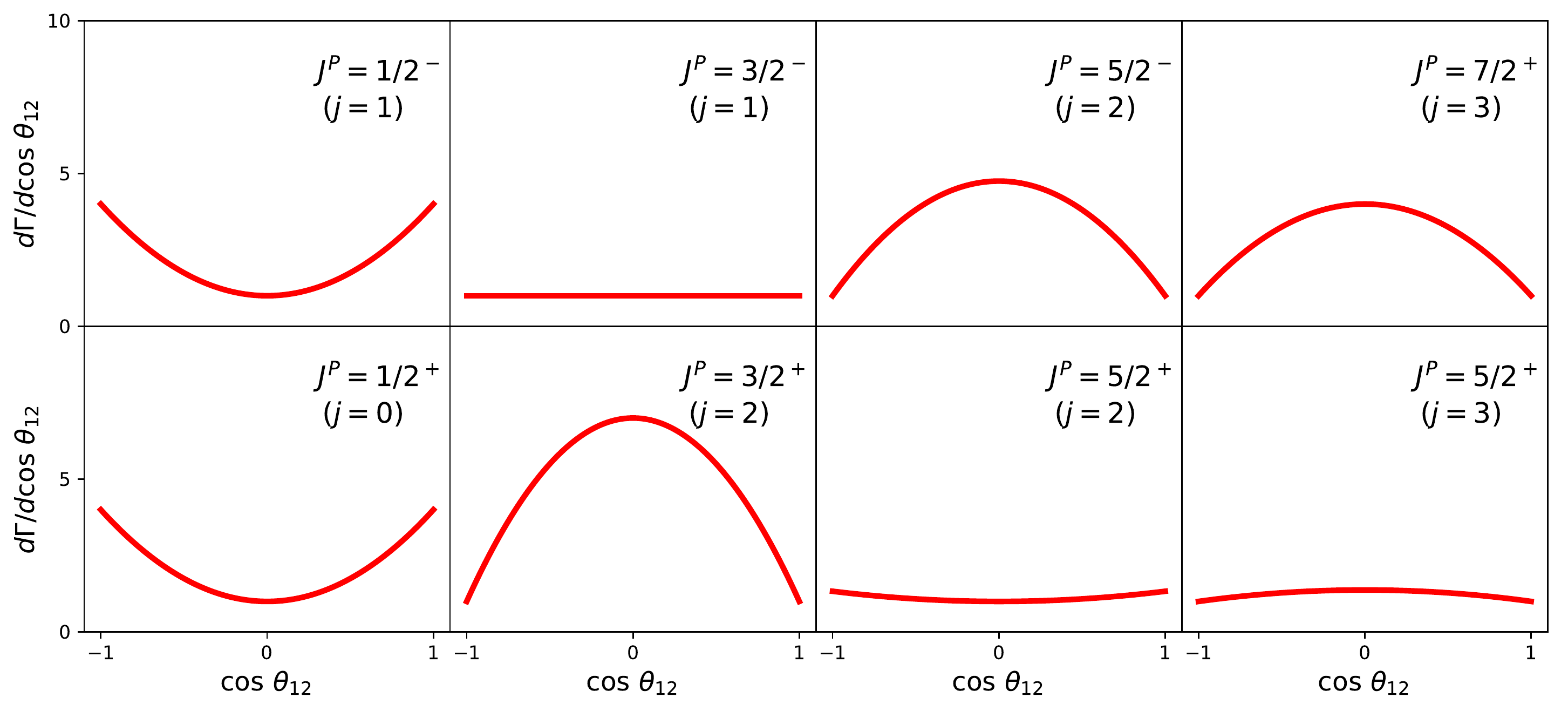}
\caption{\label{band_angle} The typical angular correlations along $\Sigma_c^{*}$ resonance band where we consider only one of the $\Sigma_c^*$ resonances appearing in the left diagram of Fig.~\ref{3body} and neglect any interference terms. The spin and parity $J^P$ of $\Lambda_c^*(2765)$ along with the brown muck spin $j$ are indicated in each figure.}
\end{figure*}

In Fig. ~\ref{band_angle}, we plot the angular correlations $W(\theta_{12})$ as functions of $\theta_{12}$ by considering only one of $\Sigma_c^*$ resonances appearing in the left diagram of Fig.~\ref{3body} for various spin and parity assignments for $\Lambda_c^*(2765)$.
The angular correlations are computed by normalizing $A_{1/2}$ equal to one,
\begin{eqnarray}
W(\theta_{12}) \propto 1 \times (1 +3 \cos^2\theta_{12}) + \tilde{R} \times 3\sin^2\theta_{12}. \quad \quad \label{sin}
\end{eqnarray}
where the ratio $\tilde{R}$ is defined by
\begin{eqnarray}
\tilde{R} = \frac{  \left|A_{3/2}(\Lambda_c^* \to \Sigma_c^*\pi)\right|^2 }{ \left|A_{1/2}(\Lambda_c^* \to \Sigma_c^*\pi)\right|^2} = \frac{|(J\ \tfrac{3}{2}\ L\ 0\ | \tfrac{3}{2}\ \tfrac{3}{2})|^2}{|(J\ \tfrac{1}{2}\ L\ 0\ | \tfrac{3}{2}\ \tfrac{1}{2})|^2}, \quad
\end{eqnarray}
with $J$ the spin of $\Lambda_c^*(2765)$ and $L$ the relative angular momentum of $\pi\Sigma_c^*$.
The ratio $\tilde{R}$ and the resulting $W(\theta_{12})$ are summarized in Table~\ref{corr}.
The Clebsh-Gordan coefficients completely determine this ratio $\tilde{R}$.
Therefore, the angular correlation can be used to determine the spin of $\Lambda_c^*(2765)$ in a model-independent way.

\begin{table}[b]
\caption{ Angular correlations along $\Sigma_c^{*}$ resonance band denoted by $W(\theta_{12})$ with various spins and parities of $\Lambda_c^*(2765)$. 
The relative angular momentum of $\pi\Sigma_c^*$ is denoted by $L$ where the forbidden one is indicated as $\cancel{L}$.
The ratio $\tilde{R}$ is defined by $\tilde{R} =|A_{3/2}(\Lambda_c^* \to \Sigma_c^*\pi)|^2/ |A_{1/2}(\Lambda_c^* \to \Sigma_c^*\pi)|^2$.
We also list the ratio $R= \Gamma(\Lambda_c^* \to \Sigma_c^*\pi)/ \Gamma(\Lambda_c^* \to \Sigma_c\pi)$ from Table~\ref{result_qm} for completeness.}
\centering
\begin{ruledtabular}
\begin{tabular}{ccccl}
   $J(j)^P$  		& 	$L$				&	$\tilde{R}$		& $R$		&  $W(\theta_{12})$	\\ \hline
  $1/2(0)^-$		& 	$\cancel{d}$ 		&	-			& -			& - \\ 
  $1/2(1)^-$		& 	$d$ 				&	0			& 0.05		& $1 +3 \cos^2 \theta_{12}$ \\
  $1/2(0)^+$		& 	$p$ 				&	0			& 0.80		& $1 +3 \cos^2 \theta_{12}$ \\
  $1/2(1)^+$		& 	$p$ 				&	0			& 0.20		& $1 +3 \cos^2 \theta_{12}$ \\ 
 \hline
  $3/2(1)^-$		& 	$s,d$ 			&	1			& 6.70		& $1$ \\
  $3/2(2)^-$		& 	$\cancel{s},d$ 		&	1			& 0.22		& $1$ \\
  $3/2(1)^+$		& 	$p,\cancel{f}$ 		&	9			& 1.99		& $1 +6 \sin^2 \theta_{12}$ \\
  $3/2(2)^+$		& 	$p,f$ 			&	9			& 0.07		& $1 +6 \sin^2 \theta_{12}$ \\ 
 \hline
  $5/2(2)^-$		& 	$d,\cancel{g}$ 		&	6			& 0.76		& $ 1 +(15/4) \sin^2 \theta_{12}$ \\
  $5/2(2)^+$		& 	$p,f $ 			&	2/3			& 13.3		& $ 1 + (1/3) \cos^2 \theta_{12}$ \\
  $5/2(3)^+$		&	$\cancel{p},f$ 		&	3/2			& 0.15		& $ 1 + (3/8) \sin^2 \theta_{12}$ \\ \hline
  $7/2(3)^+$		& 	$f,\cancel{h}$ 		&	5			& 0.35		& $1 +3 \sin^2 \theta_{12}$ \\
\end{tabular}
\label{corr}
\end{ruledtabular}
\end{table}

Fig.~\ref{band_angle} (a) and (d) show the angular correlations for  $\Lambda_c^*(2765)$ with spin 1/2 proportional to $1 +3 \cos^2\theta_{12}$ with a concave structure.
Moreover, for the case of $J^P=1/2^+$ with different brown muck spin $j$, the angular correlation also shows a concave structure as depicted in the Dalitz plot in Fig.~\ref{12p_lamrho} (g).
Since both positive and negative parity assignments to $\Lambda_c^*(2765)$ give a similar structure,
the ratio $R$, as discussed in the previous section, helps to differentiate the parities of states with the same spin.
For the higher spin states of $\Lambda_c^*(2765)$, the helicity 3/2 component has a considerable contribution, turning on the $\sin^2\theta_{12}$ dependence as described in Eq.~(\ref{sin}).
If $A_{1/2}$ and $A_{3/2}$ amplitudes are equal, the $\sin^2\theta_{12}$ dependence will cancel out the $\cos^2\theta_{12}$ dependence so that the angular correlation would be flat.
This happens only when $\Lambda_c^*\to \Sigma_c^*\pi$ decays in $s$ wave, namely for the case of $\Lambda_c^*(3/2)^-$.
For other cases, the angular correlations exhibit rather flat or convex structures depending on the value of $\tilde{R}$.
As we have discussed in section II.B, there are several cases where brown muck selection rules apply. 
For example, for  $\Lambda_c^*(5/2(3)^+)$, the $p$-wave decay into $\pi \Sigma_c^*$ is forbidden.
In this case, $f$-wave is dominant and the angular correlation changes from a concave structure $\sin^2\theta_{12}$ of $\Lambda_c^*(5/2(2)^+)$ to a convex structure $\cos^2\theta_{12}$ as shown in Fig.~\ref{band_angle}, though their angular dependence is rather weak. 

So far, we have looked at the angular correlations along one of the $\Sigma_c^*$ resonances.
In fact, there is an interference between $\Sigma_c^{*0}$ and $\Sigma_c^{*++}$ as shown in Dalitz plots in Fig.~\ref{resband}.
Therefore, the angular correlations along $\Sigma_c^{*}$ will be contaminated due to the interference, especially near $\cos\theta_{12}=-1$.
Note that the interference occurs only in the narrow region of the initial mass of $\Lambda_c^*(2765)$. 
For instance, if we plot the angular correlation at initial mass 2780 MeV or above, the interference effect is no longer significant as there are no overlapping resonance bands.
In this case, the angular correlation can be seen more clearly without significant contaminations.

\begin{figure}[t]
\centering
\includegraphics[scale=0.7]{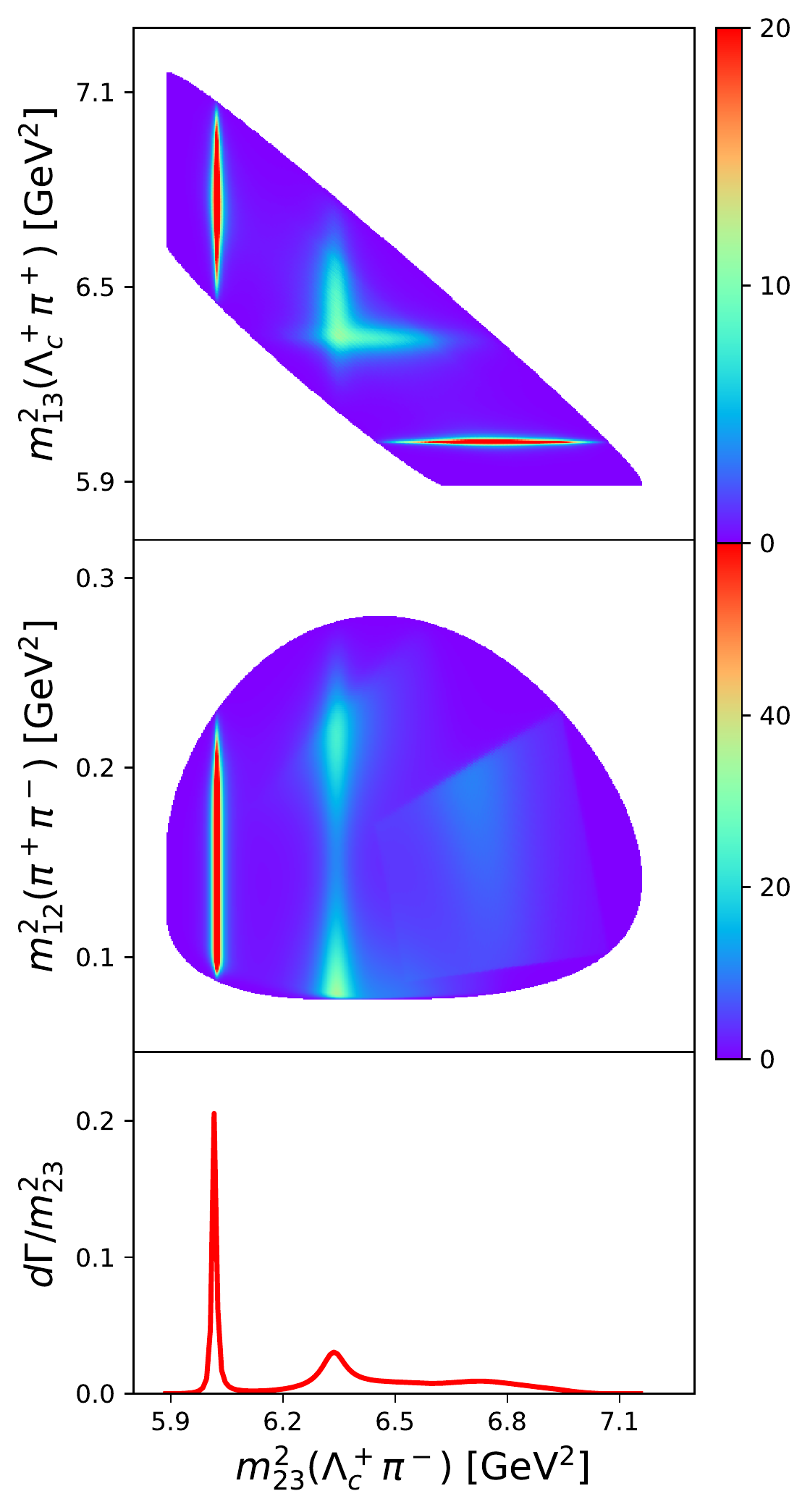}
\caption{\label{conv_dalitz} The convoluted Dalitz and invariant mass plots for $\Lambda_c^*(2765)$ with $J(j)^P=1/2(0)^+$.}
\end{figure}

\begin{figure}[t]
\centering
\includegraphics[scale=0.6]{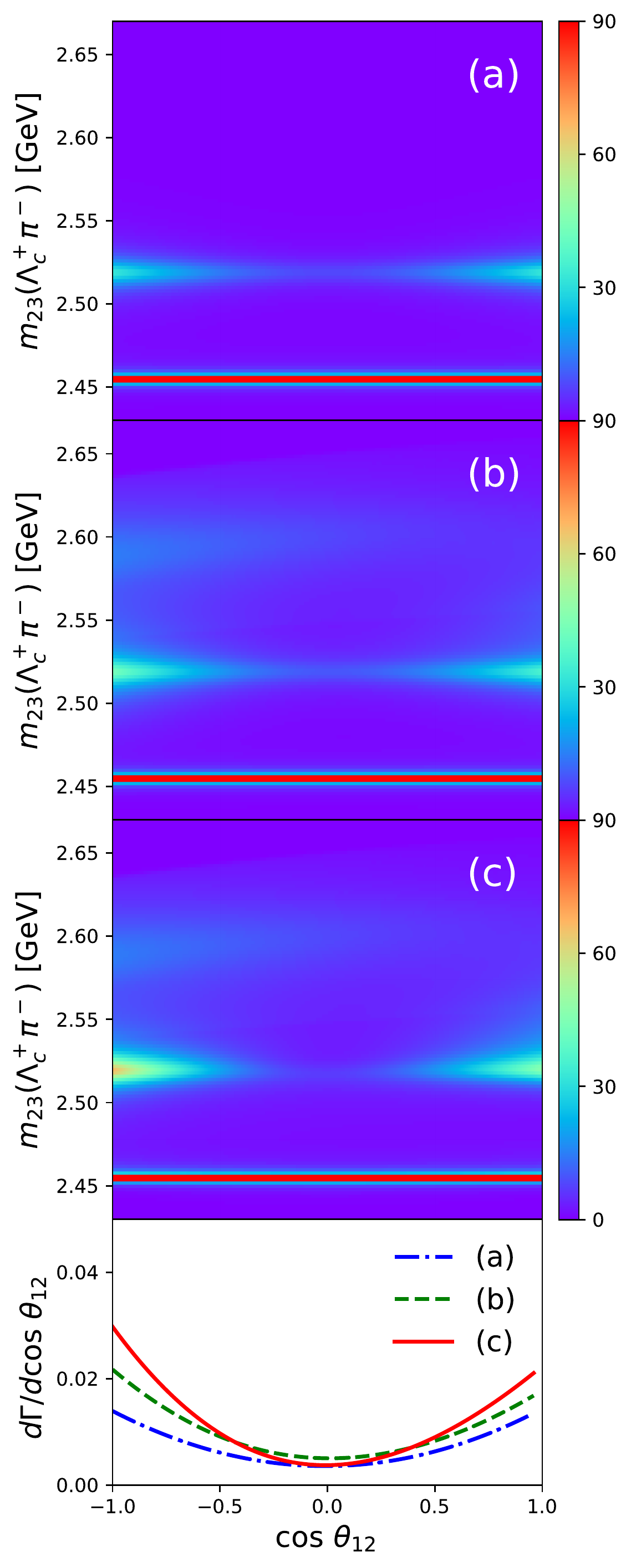}
\caption{\label{conv_correlation} The convoluted square Dalitz plots for $\Lambda_c^*(2765)$ with $J(j)^P=1/2(0)^+$ which consider (a) only $\Sigma_c^{(*)0}$, (b) $\Sigma_c^{(*)0}$ and $\Sigma_c^{(*)++}$, and (c) total amplitudes including interference terms. Note that we do not include any interference terms for (a) and (b).
Their corresponding angular correlations along $\Sigma_c^{*0}$ with a mass cut $M_{\Sigma_c^{*0}} \pm \Gamma_{\Sigma_c^{*0}}$ are given in the bottom panel.}
\end{figure}

\subsection{Effects of the finite width}

So far, all of the Dalitz plots and other observables are obtained by choosing a fixed value of the initial mass.
It is a good approximation for a narrow resonance such as $\Lambda_c^*(2625)$ with $\Gamma < 0.97$ MeV.
However, $\Lambda_c^*(2765)$ is a broad resonance with $\Gamma_{\rm exp}\approx 50$ MeV. 
Hence, a convolution is needed to directly compare theoretical results with experimental data that integrate signals over a finite mass range.
To perform a convolution, we use a Breit-Wigner form to model the mass distribution of $\Lambda_c^*(2675)$; 
\begin{eqnarray}
\tilde{\Gamma}  = \frac{1}{N} \int  \frac{\Gamma(\tilde{M}_{\Lambda_c^*})\ {\rm d}\tilde{M}_{\Lambda_c^*}}{(\tilde{M}_{\Lambda_c^*} - M_{\Lambda_c^*} )^2 + \Gamma_{\Lambda_c^*}^2/4},
\end{eqnarray}
where $\Gamma(\tilde{M}_{\Lambda_c^*})$ is the calculated decay width of $\Lambda_c^*(2765)$ which depends on the mass $\tilde{M}_{\Lambda_c^*}$. The normalization factor $N$ is defined by
\begin{eqnarray}
N =  \int \frac{{\rm d}\tilde{M}_{\Lambda_c^*}}{(\tilde{M}_{\Lambda_c^*} - M_{\Lambda_c^*} )^2 + \Gamma_{\Lambda_c^*}^2/4}.
\end{eqnarray}
We have used PDG values for the mass and width of $\Lambda_c^*(2765)$ denoted by $M_{\Lambda_c^*}$ and $\Gamma_{\Lambda_c^*}$, respectively.

To see the effect of the convolution, we show as an example of the Dalitz plot for $\Lambda_c^*(1/2^+)$ with $j=0$ in Fig.~\ref{conv_dalitz}.
In the ($m_{23}^2,m_{13}^2$) plane, four resonance bands of $\Sigma_c$ and $\Sigma_c^*$ are commonly observed. 
On the other hand, $\Sigma_c^{++}$ and $\Sigma_c^{*++}$ resonance bands are smeared out leaving two resonance bands corresponding to $\Sigma_c^0$ (left) and $\Sigma_c^{*0}$ (right) in ($m_{23}^2,m_{12}^2$) plane.
In the invariant mass plot, the peaks due to kinematical reflections disappear, as shown in the bottom panel of Fig.~\ref{conv_dalitz}.

To discuss the angular correlation along $\Sigma_c^{*0}$ resonance band, one needs to transform the convoluted Dalitz plots in Fig.~\ref{conv_dalitz} into a so-called square Dalitz plot, which is a two-dimensional plot as a function of $\cos \theta_{12}$ and $m_{23}$ as shown in Fig~\ref{conv_correlation}.
In the convoluted square Dalitz plot, the angular correlations can be seen clearly because the $\Sigma_c^{*0}$ resonance band is always spanned from $\cos \theta_{12}=-1$ to $\cos \theta_{12}=+1$ for each plot with a fixed initial mass.
If we make a narrow cut around $\Sigma_c^{*0}$, $i.e$. $M_{\Sigma_c^{*0}} \pm \Gamma_{\Sigma_c^{*0}}$, and fit the angular correlations with a polynomial of $\cos\theta_{12}$, we obtain
\begin{eqnarray}
W_a(\theta_{12}) &\propto& 1 + 2.9 \cos^2\theta_{12}, \\
W_b(\theta_{12}) &\propto& 1 + 2.9 \cos^2\theta_{12} - 0.3\cos\theta_{12},\\
W_c(\theta_{12}) &\propto& 1 + 6.0 \cos^2\theta_{12} - 0.5\cos\theta_{12}, \label{cc}\quad\quad
\end{eqnarray}
where subscripts $a, b,$ and $c$ in $W(\theta_{12})$ correspond to those labels in Fig.~\ref{conv_correlation}.
If we neglect other contributions but $\Sigma_c^{(*)0}$, the angular correlation is the same as tabulated in Fig~\ref{corr}.
Note that a small difference in $\cos^2\theta_{12}$ coefficient is due to $\Sigma_c^0$ contribution.
When we add other contributions from $\Sigma_c^{(*)++}$ without including interference terms, the angular correlation becomes slightly asymmetric because there is an overlap between $\Sigma_c^{*0}$ and  $\Sigma_c^{*++}$ resonances in the lower region of the upper Dalitz plot in Fig~\ref{conv_dalitz}.
Finally, if we consider the interference terms, the angular correlation considerably changes as shown in Eq.~(\ref{cc}), but it still exhibits a concave structure as seen in Fig.~\ref{conv_correlation}.
In general, the interference terms modify the angular correlations, but they do not change the characteristic shape of the angular correlations in Fig.~\ref{band_angle}.
For $\Lambda_c^*(2765)$, there is an accidental interference between $\Sigma_c^{*0}$ and $\Sigma_c^{*++}$ resonances.
However, for higher excited states of $\Lambda_c^*$ baryons, $e.g.$ $\Lambda_c^*(2880)$, those $\Sigma_c^*$ resonances are well separated such that the analysis becomes easier.

\vspace{-0.4cm}
\section{SUMMARY}

In this work, we have investigated the three-body decay of $\Lambda_c^*(2765) \to \Lambda_c^+ \pi^+ \pi^-$.
Here, we focus on the sequential processes going through $\Sigma_c^{(*)}$ resonances, by accepting that the contribution of the direct process is small through the experimental observation~\cite{Abe:2006rz}. 
The reason for the small contribution of the direct process in $\Lambda_c^*(2765)$ decay is unknown.

We have performed the Dalitz plot analysis with various spin and parity assignments of $\Lambda_c^*(2765)$.
Employing effective Lagrangians in the non-relativistic framework, we have computed all possible two-body decays of $\Lambda_c^*(2765) \to \Sigma_c^{(*)}\pi$ by means of the quark model for all possible configurations up to $2\hbar\omega$ regions.
The results are transformed into various coupling constants in the effective Lagrangians.  

It turns out that geometric and dynamical factors determine the structures of the Dalitz plots.
Geometric factors are model-independent and are characterized by the spin and parity of participating particles and underlying symmetry.
They are angular correlations that are determined by spin, and the ratios $R$ that are dominated by the parity that determines the partial wave of decaying particles.
In contrast, dynamical factors are model-dependent such as the interaction strengths and form factors.
The dynamical factor is taken into account by using the quark model as input, which characterizes the strengths of the $\Sigma_c^{(*)}$ intermediate states.

From absolute values of decay widths, one can not decide which quark model configuration is suitable for $\Lambda_c^*(2765)$.
However, the ratios $R$ are sensitive to the configurations, which are reflected in Dalitz and invariant mass plots.
Moreover, it is found that the angular correlations along the $\Sigma_c^*$ resonance band in the Dalitz plots are sensitive to the spin and parity of $\Lambda_c^*(2765)$.
Finally, we have investigated the effect of the finite width of $\Lambda_c^*(2765)$ and the interference terms.
In convoluted Dalitz plots, we have found that the kinematical reflections are smeared out, but the angular correlations can be still observed clearly.
The interference terms can contaminate the angular correlation but do not change its characteristic shape.
Therefore, the information about the ratio $R$ and the angular correlation would shed the light on the spin and parity of $\Lambda_c^*(2765)$.

A similar angular correlation analysis can also be done for three-body decays of charm-strange baryons, in particular for $\Xi_c^*(2970) \to \Xi_c \pi \pi$ decay~\cite{Lesiak:2008wz,Yelton:2016fqw}. 
In this case, there is no kinematical reflection of $\Xi_c^*(2645)$ intermediate state and $\Xi_c'$ has a negligible width, resulting no significant contaminations from interferences.
Furthermore, we can apply the analysis to the bottom sectors such as recently observed $\Lambda_b^*$ baryons in $\Lambda_b \pi \pi$ invariant mass~\cite{Aaij:2019amv,Sirunyan:2020gtz,Aaij:2020rkw} in determination of their spin and parity.
We will discuss these issues elsewhere.

\begin{acknowledgments}
We thank to Dr. Changwoo Joo for the discussion about the experimental situation.
This work is supported by a scholarship from the Ministry of Education, Culture, Science and Technology of Japan for A. J. Arifi,
and also Grants-in-Aid for Scientific Research, Grants No. 17K05443(C) for H. Nagahiro and Grants No. 17K05441(C) for A. Hosaka.
Finally, we thank support from the Reimei Research Promotion project (Japan Atomic Energy Agency) in completion of this work.
\end{acknowledgments}

\end{document}